\documentclass[%
reprint, 
prx,
amsmath,amssymb,
aps,
floatfix,
]{revtex4-2}

\usepackage{graphicx}
\usepackage{physics}
\usepackage{bm} 

\usepackage{booktabs}
\usepackage{caption}
\usepackage{subcaption}
\usepackage{makecell} 
\usepackage{xcolor} 
\usepackage{soul}
\usepackage{ulem}

\usepackage[colorlinks, allcolors=blue]{hyperref}

\def\urbana{
The Anthony J. Leggett Institute for Condensed Matter Theory and IQUIST and NCSA Center for Artificial Intelligence Innovation and Department of Physics, University of Illinois at Urbana-Champaign, IL 61801, USA}

\begin{document}
\title{Neural network backflow for ab-initio solid calculations}

\author{An-Jun Liu} 
\affiliation{\urbana}
\author{Bryan K. Clark}
\affiliation{\urbana}

\begin{abstract}
Accurately simulating extended periodic systems is a central challenge in condensed matter physics. Neural quantum states (NQS) offer expressive wavefunctions for this task but face issues with scalability.
In this work, we successfully extend the neural network backflow (NNBF) approach to \textit{ab-initio} solid-state materials. Building on our scalable optimization framework for molecules [Liu \textit{et al.}, PRB \textbf{112}, 155162 (2025)], we introduce a two-stage pruning strategy to manage the massive configuration space expansions: by utilizing a computationally cheap, physics-informed importance proxy, we devote exact NNBF amplitude evaluations solely to the most relevant determinants, significantly improving optimization efficiency, energy estimation, and convergence. Our framework achieves state-of-the-art accuracy across diverse solid-state benchmarks. For 1D hydrogen chains, NNBF matches or surpasses DMRG and AFQMC, remains robust in strongly correlated bond-breaking regimes where coupled-cluster methods fail, and smoothly extrapolates to the TDL. We further demonstrate its scalability by computing ground-state potential energy curves for 2D graphene and 3D silicon. Finally, ablation studies validate the computational savings of our pruning strategy and highlight the dependence of the NNBF energies on basis sets. 
\end{abstract}

\maketitle

\section{Introduction}

The \textit{ab-initio} computation of electronic structure in periodic materials is a central challenge in condensed matter physics and quantum chemistry, essential for predicting material properties from first principles. However, the exact solution of the many-body Schrödinger equation scales exponentially with system size, necessitating sophisticated approximations. Traditional  coupled-cluster (CC) method \cite{Coester1960}, serve as the gold standard for weakly correlated systems but suffer severe breakdowns in the presence of strong static correlation, such as during bond breaking. Conversely, while the Density Matrix Renormalization Group (DMRG) method \cite{White1992,White1999,Chan2016} excels for one-dimensional systems, it struggles to scale efficiently in higher dimensions. Phaseless Auxiliary-Field Quantum Monte Carlo (AFQMC) \cite{Blankenbecler1981,Zhang1997,Zhang2003} is often accurate, but doesn't provide a variational upper bound and lacks a compact, closed-form variational wave function, making the subsequent evaluation of arbitrary observables computationally demanding.

Recently, Neural Quantum States (NQS) have emerged as a highly promising alternative, leveraging the universal approximation capabilities of deep neural networks to compactly encode complex, highly entangled many-body wave functions. 
However, the vast majority of NQS applications have been heavily restricted to lattice or molecular systems\cite{Liu2024,Liu2025,Zhao2023,Li2023,Shang2023,Wu2023,Malyshev2023,Liu2024,Li2024,Malyshev2024,Knitter2024,Pfau2020,Hermann2020,Di2019,Loehr20025,RobledoMoreno2022,zejun2023}.
Applications of NQS to \textit{ab-initio} solid-state materials in both first and second quantization remain remarkably sparse \cite{Shang2024-DMET,Shang2024-Solid,Yoshioka2021,DeepSolid}.  
The Neural Network Backflow (NNBF) \cite{Di2019,zejun2023,Liu2024,Liu2025,Loehr20025,Zhuo2022} ansatz has demonstrated state-of-the-art accuracy in molecular systems and first-quantized solids but has not yet been applied to ab-initio materials. 
In this work, we address this omission by extending our recent, highly scalable ab-intio NNBF framework—originally developed for molecular systems (Ref.~\onlinecite{Liu2025})—to periodic solid-state materials. Our previous work overcame the difficulties of sampling highly peaked distributions and the prohibitive quartic scaling of local energy evaluations through a unified approach: periodically constructing a compact yet important subspace, reusing pre-computed information for truncated local energy evaluations, and employing an improved stochastic sampling method for unbiased energy estimation. To extend this foundation to periodic solid materials, we introduce an efficient two-stage pruning strategy.
In the first stage, a computationally cheap, physics-informed importance proxy filters the massively expanded connected space into a highly representative intermediate pool. In the second stage, exact NNBF amplitude evaluations are performed solely on this reduced pool to construct the final, optimal target space $\mathcal{U}$. This algorithmic innovation drastically reduces computational overhead while successfully capturing the most physically relevant configurations while optimizing the wave-function, allowing the NNBF ansatz to efficiently and accurately solve complex second-quantized \textit{ab-initio} solids.

We demonstrate the efficacy of this extended framework across a variety of periodic benchmarks. First, we compute the potential energy curves of 1D hydrogen chains under both open and periodic boundary conditions, successfully extrapolating our results to the thermodynamic limit. In these systems, our method matches or exceeds the performance of AFQMC, DMRG, and conventional CC methods, showing particular robustness in the strongly correlated bond-breaking regime where perturbation-based methods fail. We then showcase the method's scalability to higher-dimensional materials by computing the ground-state potential energy curves for 2D hexagonal graphene and 3D face-centered cubic silicon. Finally, through extensive ablation studies, we quantitatively validate the computational efficiency of our two-stage proxy-based pruning strategy and highlight the critical role of the single-particle basis set for maintaining NQS accuracy in strongly correlated regimes.

\section{Methods}

In this section, we first present an overview of the NNBF background, summarize the prior optimization framework we build upon, and detail the algorithmic improvements introduced in this study to extend the method to solid-state systems.

\subsection{VMC with NNBF}

Neural Quantum States (NQS), and specifically the Neural Network Backflow (NNBF) architecture, have proven highly effective for representing fermionic wavefunctions in second quantization \cite{Di2019, zejun2023, Zhuo2022, Liu2024, Liu2025, Loehr20025}. For a many-electron system defined by $N_e$ electrons and $N_o$ single-particle orbitals (SPOs), the generic electronic Hamiltonian takes the form:
\begin{equation}
\hat{H} = \sum_{ij\sigma}t_{ij}\hat{c}_{i\sigma}^\dagger \hat{c}_{j\sigma} + \frac{1}{2} \sum_{ijkl\sigma\sigma'}V_{ijkl}\hat{c}_{i\sigma}^\dagger  \hat{c}_{j\sigma'}^\dagger \hat{c}_{l\sigma'} \hat{c}_{k\sigma}
\end{equation}
where the indices $i,j,k,l$ iterate over the $N_o$ SPOs and $\sigma, \sigma'$ denote spin. The corresponding many-electron wavefunction is expressed as $\ket{\psi} = \sum_{i}\psi(\mathbf{x}_i)\ket{\mathbf{x}_i}$, where $\ket{\mathbf{x}_i}=\ket{x_i^{1\uparrow},\dots,x_i^{N_o\uparrow},x_i^{1\downarrow},\dots,x_i^{N_o\downarrow}}$ is the $i$-th computational basis vector, and $x_i^j \in \{0,1\}$ denotes the occupation of the $j$-th spin-orbital.

To approximate this state, a generic NNBF wavefunction is defined as:
\begin{equation}
\psi_\theta(\mathbf{x}_i)=\sum_{m=1}^{D}\det[\Phi^m_{j=\{l | x_i^l=1\},k}(\mathbf{x}_i;\theta)]
\end{equation}
where $\Phi^m_{jk}$ are the ``configuration-dependent'' spin-orbitals (neural orbitals) output by the internal neural network, $D$ is the number of determinants, and $\theta$ represents the model parameters.

To optimize the NNBF ansatz to approximate the ground state, training relies on gradient-based minimization of the variational energy:
\begin{equation}
E_\theta=\frac{\bra{\psi_\theta}\hat{H}\ket{\psi_\theta}}{\braket{\psi_\theta}{\psi_\theta}}=\mathbb{E}_{p_{\theta}(\mathbf{x})} \left[E_{l}(\mathbf{x})\right]
\end{equation}
where the local energy is $E_{l}(\mathbf{x}) = \frac{\bra{\psi_\theta}\hat{H}\ket{\mathbf{x}}}{\braket{\psi_\theta}{\mathbf{x}}}$. The exact gradient with respect to the network parameters is:
\begin{equation}
\nabla_\theta E_\theta = 2\Re{ \mathbb{E}_{p_{\theta}(\mathbf{x})}\left[ \frac{\partial \ln{\abs{\psi_\theta(\mathbf{x})}}}{\partial \theta} \left[E_{l}(\mathbf{x}) - E_\theta \right] \right] }
\end{equation}

\subsection{Background}\label{sec:background}

Optimizing molecular Hamiltonians with NQS generally faces two severe bottlenecks. First, the probability distribution is highly peaked around the Hartree–Fock (HF) state and nearby excited states \cite{Bytautas2009, Anderson2018}. This causes standard Markov Chain Monte Carlo (MCMC) methods to waste computational resources excessively resampling dominant configurations. Second, the local energy evaluation scales quartically with system size, making exact computations prohibitive for larger systems.

Ref.~\onlinecite{Liu2025} resolved these issues by introducing a scalable optimization framework. At a given training step, a connected space $\mathcal{C}$ is expanded from a small core space $\mathcal{V}$ of unique dominant configurations. The amplitudes for all the configurations inside $\mathcal{V}\cup\mathcal{C}$ are computed, and the top $|\mathcal{V}|N_{conn}/l$ unique configurations with the largest amplitude moduli are selected to form a compact target space $\mathcal{U}$, where $N_{conn}$ is the number of connected terms and $l$
is a predefined speedup factor.
This target space $\mathcal{U}$ is fixed for the subsequent $l-1$ steps, and only the amplitudes inside it will be computed.

To construct unbiased estimators for the energy and its gradient, a Gumbel top-$k$ trick is employed to sample a set $\mathcal{S}$ from $\mathcal{U}$, assigning an importance weight $w_i(\kappa)$ to each sample $\ket{\mathbf{x}_i}\in\mathcal{S}$. The local energy calculation is correspondingly streamlined by truncating the sum to include only terms within $\mathcal{U}$, denoted as $E_l(x_i|\mathcal{U})$. Together, $w_i(\kappa)$ and $E_l(x_i|\mathcal{U})$ are used to estimate the objective functions efficiently before applying parameter updates via AdamW \cite{AdamW}. Finally, to ensure robust exploration, a concurrent set of $|\mathcal{C}|N_e$ MCMC walkers is injected at the end of the $l$-step interval to update the core space $\mathcal{V}$. This framework drastically boosts training efficiency without sacrificing accuracy, allowing NNBF to achieve results competitive with state-of-the-art ab-initio methods like HCI \cite{Holmes2016,Sharma2017}, ASCI \cite{Tubman2016,Tubman2018}, FCIQMC \cite{Cleland2012}, and DMRG \cite{Chan2004}.

\subsection{Algorithmic improvements}\label{sec:algorithmic_improvement}

In this work, we extend the framework of Ref.~\onlinecite{Liu2025} to periodic solid-state systems. Unlike molecular systems where wavefunctions are typically expanded in atomic basis sets $\{\chi_i\}$, crystalline states are instead expanded in Bloch atomic orbitals $\phi_{i\mathbf{k}}(\mathbf{r}) = \frac{1}{\sqrt{N}} \sum_{n} e^{i\mathbf{k}\cdot\mathbf{R}_n} \chi_i(\mathbf{r} - \mathbf{R}_n)$ to embed the system's periodicity, in which $\mathbf{R}_n$ is a lattice vector, $\mathbf{k}$ is a Bloch momentum sampled in the first Brillouin zone, and $N$ is the number of unit cells in the crystal. By performing a periodic Hartree–Fock calculation using PySCF \cite{pyscf}—where Coulomb singularities are treated via the Ewald method—we obtain a basis of crystalline HF orbitals. The $i$-th state at $\mathbf{k}$ is expanded as $\varphi_{i\mathbf{k}}(\mathbf{r}) = \sum_j C_{j i}(\mathbf{k}) \phi_{j\mathbf{k}}(\mathbf{r})$, where $C_{ji}(\mathbf{k})$ are the coefficients determined by this procedure.

Note that unlike orbitals in quantum chemistry, the overlaps between Bloch states are highly non-local due to the periodicity of the system. This non-locality is directly reflected by the structure of the Coulomb tensor in the resulting second-quantized crystalline Hamiltonian:
\begin{equation}
\begin{split}
    \hat{H} &= \sum_{ij\mathbf{k}\sigma} h_{ij}^\mathbf{k} \hat{c}_{i\mathbf{k}\sigma}^\dagger \hat{c}_{j\mathbf{k}\sigma} \\
    &\quad + \frac{1}{2} \sum_{\substack{ijkl \\ \mathbf{k}_1\mathbf{k}_2\mathbf{k}_3\mathbf{k}_4 \\ \sigma\sigma'}} V_{ijkl}^{\mathbf{k}_1\mathbf{k}_2\mathbf{k}_3\mathbf{k}_4} \hat{c}_{i\mathbf{k}_1\sigma}^\dagger \hat{c}_{j\mathbf{k}_2\sigma'}^\dagger \hat{c}_{l\mathbf{k}_4\sigma'} \hat{c}_{k\mathbf{k}_3\sigma}.
\end{split}
\end{equation}
The momentum sum is evaluated over a uniform grid in the first Brillouin zone. Because the two-electron integral is only nonzero when crystal momentum is conserved ($\mathbf{k}_1 + \mathbf{k}_2 - \mathbf{k}_3 - \mathbf{k}_4 = \mathbf{G}$, where $\mathbf{G}$ is a reciprocal lattice vector), the total number of terms scales as $\mathcal{O}(N_o^4N_k^3)$. In this work, we restrict our focus to $\Gamma$-point calculations ($N_k=1$), ensuring all Hamiltonian matrix elements are real but work in a supercell that consists of many unit cells.

\begin{figure}[htbp!]
    \centering
    \includegraphics[width=0.85\linewidth]{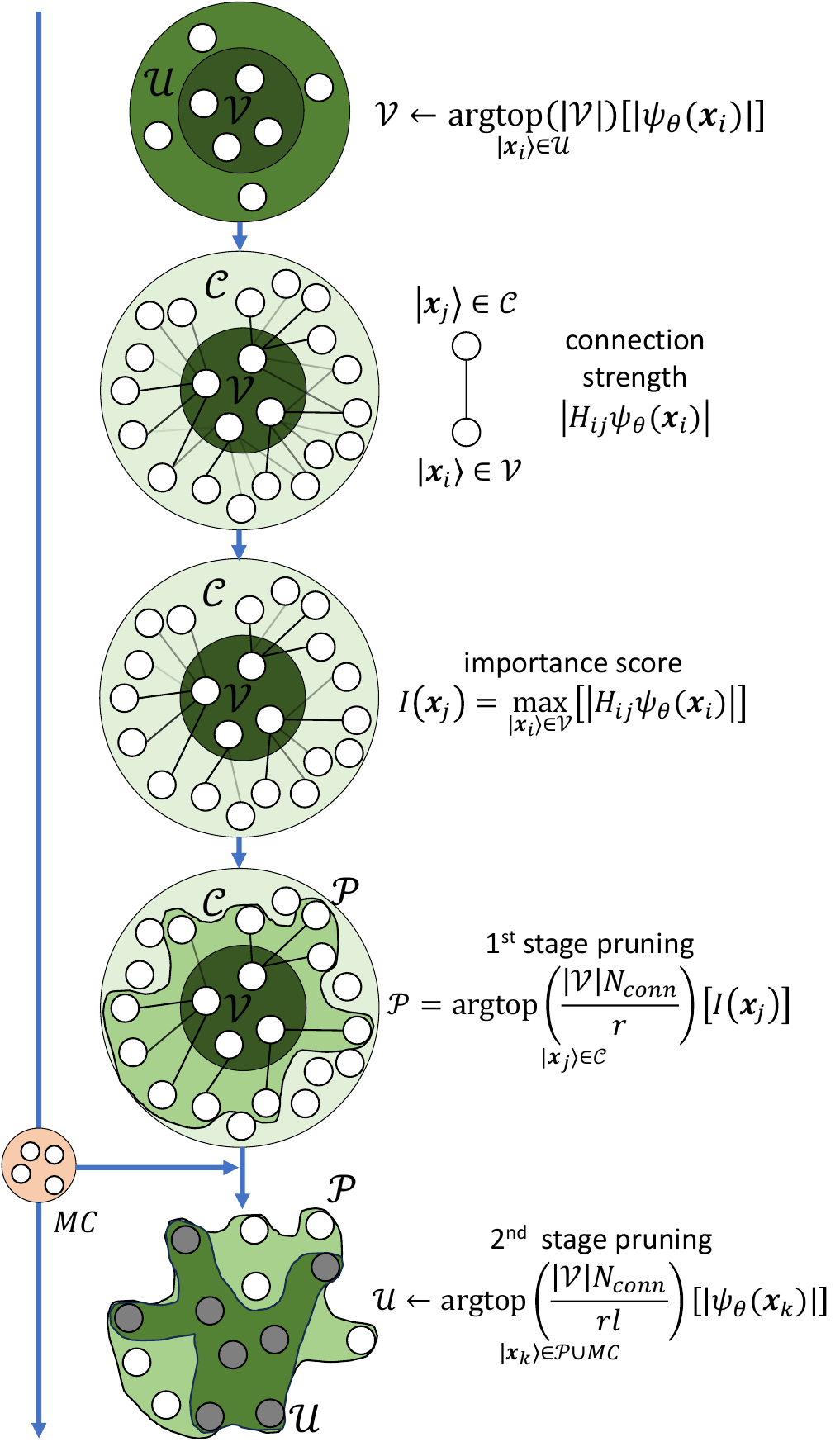}
    \caption{Schematic representation of the two-stage pruning algorithm. \textbf{Circle 1:} The core space $\mathcal{V}$ consists of the $\abs{\mathcal{V}}$ configurations with the largest amplitude moduli from the previous target space $\mathcal{U}$. \textbf{Circle 2:} A connected space $\mathcal{C}$ is expanded from $\mathcal{V}$ via non-zero Hamiltonian matrix elements. Each connected configuration $\ket{\mathbf{x}_j}\in\mathcal{C}$ may link to multiple core configurations $\ket{\mathbf{x}_i}\in\mathcal{V}$, and these connection strengths are quantified as $\abs{H_{ij}\psi_\theta(\mathbf{x}_i)}$. \textbf{Circle 3:} The importance score of each connected configuration $\ket{\mathbf{x}_j}$ is then defined as its maximum connection strength: $I(\mathbf{x}_j) = \max_{\ket{\mathbf{x}_i} \in \mathcal{V}} \left| \psi_\theta(\mathbf{x}_i) H_{ij} \right|$. \textbf{Circle 4:} The first stage of pruning selects the $\abs{\mathcal{V}}N_{conn}/r$ unique connected configurations with the largest importance scores to form an intermediate pool space $\mathcal{P}$. The predefined reduction factor r ensures that the size of $\mathcal{P}$ matches the number of configurations that would be generated by fully expanding a reduced core space of size $\abs{\mathcal{V}}/r$. \textbf{Circle 5:} The second stage of pruning calculates the exact NNBF amplitudes for the combined space $\mathcal{P}\cup\mathcal{MC}$, where $\mathcal{MC}$ represents configurations proposed by persistent MCMC walkers running concurrently to provide stochastic exploration. Finally, the $\abs{\mathcal{V}}N_{conn}/rl$ elements with the largest amplitude moduli are selected to form the new target space $\mathcal{U}$. This space $\mathcal{U}$ is fixed and used for the subsequent $l$ steps, where $l$ is a predefined speedup factor.}
    \label{fig:algorithmic_improvements}
\end{figure}

We introduce two primary algorithmic modifications to adapt the previous framework for solids efficiently. The first addresses a key inefficiency: the connected space $\mathcal{C}$ expanded from a core space $\mathcal{V}$ may contain redundant configurations (for example, the single excitations of a core configuration, which itself is a single excitation relative to another core configuration, are part of the double excitations of the latter core configuration), particularly when spin-flip symmetry is enforced. To resolve this, we introduce a more efficient two-stage pruning algorithm to construct the target space $\mathcal{U}$, as visualized in Fig.~\ref{fig:algorithmic_improvements}.

First, the $\abs{\mathcal{V}}$ unique configurations with the largest amplitude moduli are selected from $\mathcal{U}$ to form the core space $\mathcal{V}$, and a connected space $\mathcal{C}$ is expanded from $\mathcal{V}$ via Hamiltonian matrix elements. Each connected configuration $\ket{\mathbf{x}_j}\in\mathcal{C}$ links to multiple core configurations $\ket{\mathbf{x}_i}\in\mathcal{V}$. We quantify the strength of these links by $|\psi_\theta(\mathbf{x}_i)H_{ij}|$, which captures both strong Hamiltonian coupling and proximity to dominant configurations. The maximum connection strength defines the importance score for $\ket{\mathbf{x}_j}$:
\begin{equation}\label{eq:importance_score}
I(\mathbf{x}_j) = \max_{\ket{\mathbf{x}_i} \in \mathcal{V}} \left| \psi_\theta(\mathbf{x}_i) H_{ij} \right|.
\end{equation}
This metric parallels the deterministic selection heuristic used in Heat-bath Configuration Interaction (HCI) \cite{Holmes2016}. 

In the first pruning stage, we select the $|\mathcal{V}|N_{conn}/r$ unique elements with the largest importance scores to form an (ideally) deduplicated pool space $\mathcal{P}$. Crucially, this step does not require any NNBF amplitude evaluations, and $r$ is a predefined reduction factor, meaning the size of $\mathcal{P}$ matches the effective size of a connected space that would be expanded from a smaller core space of size $\abs{\mathcal{V}}/r$ without any pruning.

In the second stage, we evaluate the exact amplitudes for the configurations in $\mathcal{P}$ and select the top $(|\mathcal{V}|N_{conn}/r)/l$ elements with the largest amplitude moduli to form the final, highly dominant target space $\mathcal{U}$. This proxy-based pre-filtering creates a superior target space, yields more accurate gradient estimations, and ultimately leads to better training results (their effects are investigated in detail in Sec.~\ref{sec:ablation_study}). A schematic overview of this two-stage pruning algorithm is presented in Fig.~\ref{fig:algorithmic_improvements}.

The second modification refines the MCMC injection strategy. Injecting MCMC walkers directly into the core space $\mathcal{V}$ can be suboptimal; walkers are often populated by dominant configurations already present in $\mathcal{V}$, while semi-important configurations proposed by the walkers are discarded before their exact amplitudes can be evaluated. Instead, we now inject the MCMC walkers into the formation of the pool space $\mathcal{P}$. This allows the exact NNBF amplitude evaluation—the most precise metric available—to dictate which of these proposed configurations ultimately survive the trim into $\mathcal{U}$. Furthermore, because we maintain persistent MCMC walkers without restarting them at each step, long equilibration times are unnecessary. 
We perform just 1 proposal move per training step using $0.1|\mathcal{U}|$ walkers, effectively hiding the sampling cost within the main amplitude evaluation subroutine.

\section{Results}

We first evaluate the performance of the NNBF ansatz, combined with the algorithmic improvements proposed in Sec.~\ref{sec:algorithmic_improvement}, on a 1D linear hydrogen chain. We begin by presenting the STO-6G dissociation curve for a finite chain of ten hydrogen atoms under open boundary conditions, which has been used as the benchmark system for various many-body methods\cite{Motta2017}. We then demonstrate the extrapolated dissociation curve in the thermodynamic limit. Next, we transition to periodic boundary conditions, presenting the STO-6G dissociation curve for a periodic H$_{10}$ chain alongside thermodynamic limit extrapolations at a selected interatomic distance. Finally, to demonstrate the broader capability of our method, we apply it to real solid-state systems, specifically 2D graphene and 3D silicon using the GTH-DZV basis set.

To validate the algorithmic improvements introduced in Sec.~\ref{sec:algorithmic_improvement}, we perform an ablation study benchmarking the effectiveness of the proxy importance score in Eq.~\eqref{eq:importance_score} against both a random importance score and the unpruned expansion method from Ref.~\onlinecite{Liu2025}. Furthermore, we demonstrate the quality of the proxy score for capturing the true ordering of amplitude moduli. We visualize the cumulative probability captured as a function of the pool space size $\abs{\mathcal{P}}$ and compare it against both random selection and the optimal selection based on exact amplitude moduli. Lastly, we present a separate ablation study highlighting the critical role of basis set selection in NQS training. Specific details
regarding the (default) neural network architectures, hyperparameters, training protocols, and the post-training MCMC inference procedure are provided in Appendix \ref{appx:experimental_setup}.

\subsection{Linear Hydrogen Chain}

The linear hydrogen chain serves as an ideal benchmark for evaluating many-body electronic structure methods because the strength of its electron correlation can be tuned via the interatomic separation. Unlike the Hubbard model, which captures only on-site interactions, the hydrogen chain explicitly includes long-range Coulomb interactions. As a result, its ground-state properties reflect key features that are essential for generalizing to real materials.

\begin{figure}[htbp!]
    \centering
    \includegraphics[width=\linewidth]{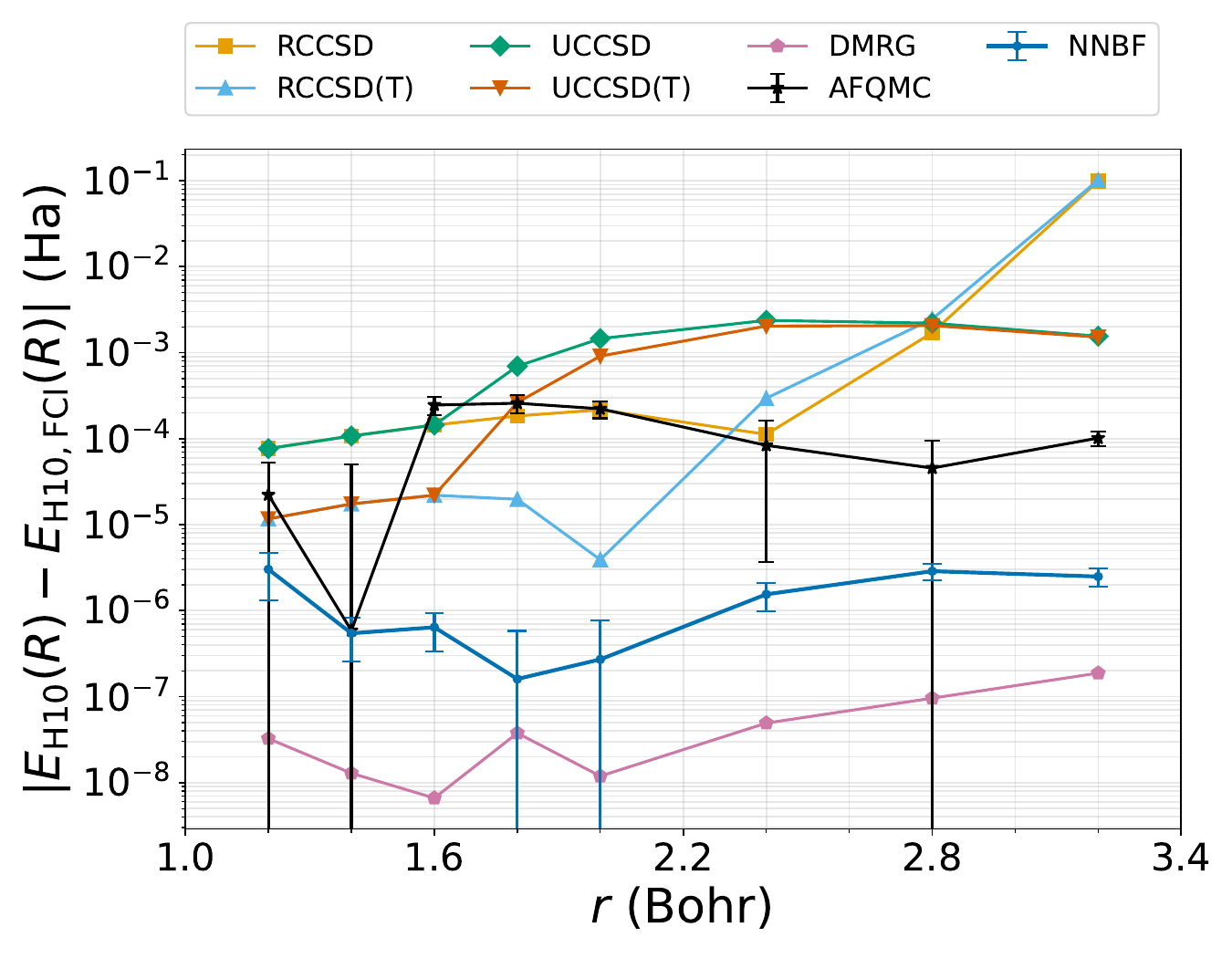}
    \caption{Potential energy curve for H$_{10}$ under OBC using the STO-6G basis set. The energy errors of our NNBF ansatz, conventional quantum chemistry methods [CCSD, CCSD(T)], AFQMC, and DMRG are plotted relative to the exact FCI reference energy. The reported NNBF energies are obtained via the following protocol: three independent training runs using split localized Pipek--Mezey (PM) molecular orbitals are performed at each atomic separation (see Appendix~\ref{appx:experimental_setup} for detailed settings), and a post-training MCMC inference is used to evaluate the variational energy of each run. The model yielding the lowest of these energies is selected. A final, independent MCMC inference is then conducted on this optimal model to obtain the unbiased energy estimate displayed in the figure. Data for all other benchmark methods are taken from Ref.~\onlinecite{Motta2017}.}
    \label{fig:H10_OBC_curve}
\end{figure}

Building on the extensive benchmarks established by Motta \textit{et al.}~\cite{Motta2017}, we first investigate the system under open boundary conditions (OBC). The ground-state energies for a 10-atom chain at various atomic separations are calculated using our NNBF ansatz and compare against traditional methods—(R/U)CCSD and (R/U)CCSD(T)—as well as state-of-the-art ab-initio approaches like AFQMC and DMRG\cite{Motta2017}. As shown by the energy errors relative to the exact FCI results in Fig.~\ref{fig:H10_OBC_curve}, NNBF matches the accuracy of AFQMC and DMRG, outperforming conventional gold-standard coupled-cluster methods. Notably, at large atomic separations where static correlation dominates, single-reference methods like RCCSD break down, whereas NNBF remains robust and reliable in this strongly correlated regime.

\begin{figure}[htbp!]
    \centering
    \includegraphics[width=\linewidth]{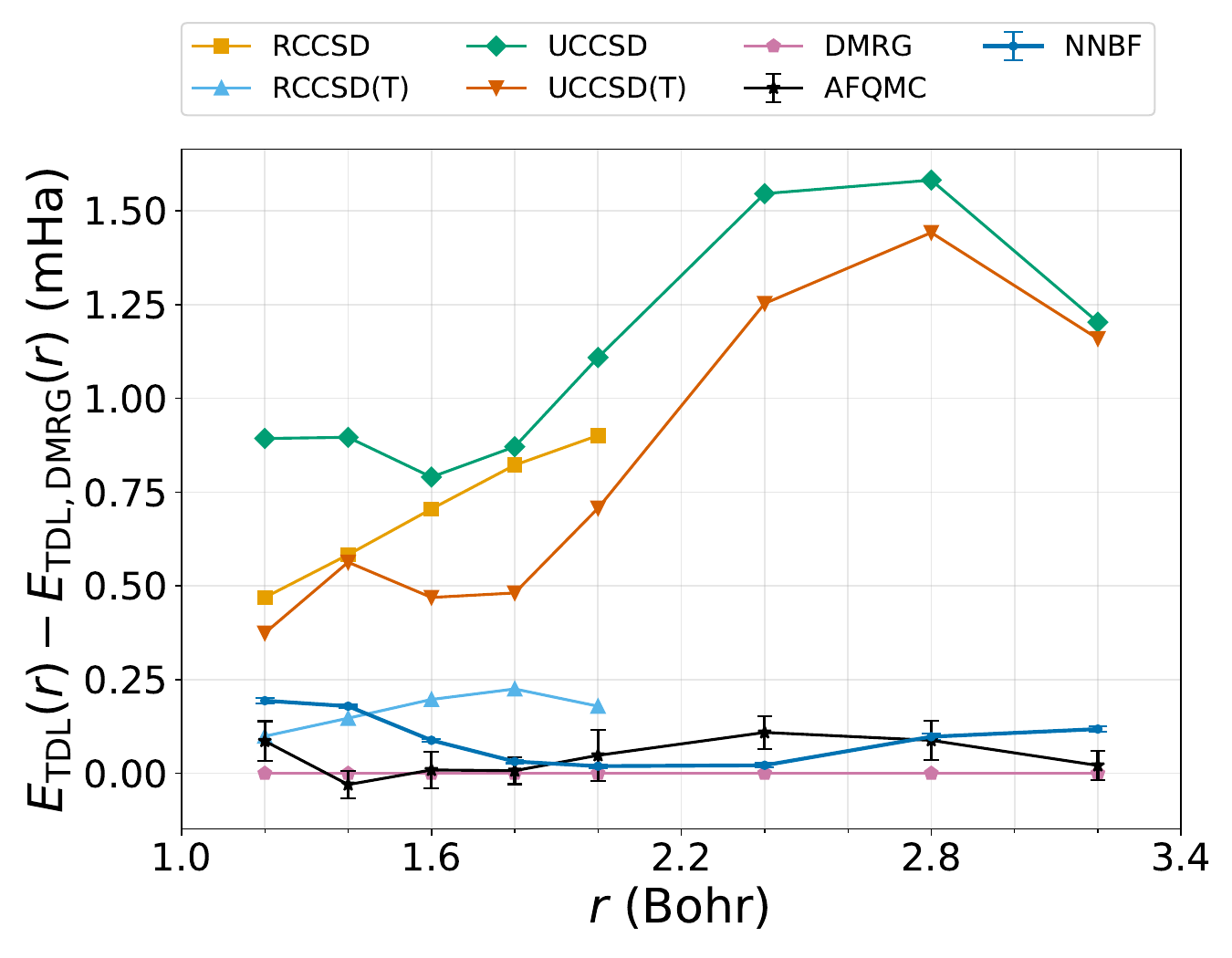}
    \caption{Potential energy curve at the thermodynamic limit (TDL) for the linear hydrogen chain under OBC using the STO-6G basis set. The TDL-extrapolated energies of our NNBF ansatz, conventional quantum chemistry methods [CCSD, CCSD(T)], and AFQMC are plotted relative to the TDL-extrapolated DMRG reference. Details of the NNBF TDL extrapolation are provided in Appendix~\ref{appx:OBC_TDL_extrapolation}. Data for all other benchmark methods are taken from Ref.~\onlinecite{Motta2017}.}
    \label{fig:OBC_TDL_curve}
\end{figure}

To eliminate finite-size effects, we extrapolate the OBC results to the thermodynamic limit (TDL), with full details of the procedure provided in Appendix~\ref{appx:OBC_TDL_extrapolation}. As illustrated in Fig.~\ref{fig:OBC_TDL_curve}, comparing our TDL extrapolation against DMRG references reveals that the algorithmically enhanced NNBF yields energies accurate to within 0.2 mHa per particle across all bond lengths. This confirms that our method's performance scales competitively alongside DMRG and AFQMC.

\begin{figure}[htbp!]
    \centering
    \includegraphics[width=\linewidth]{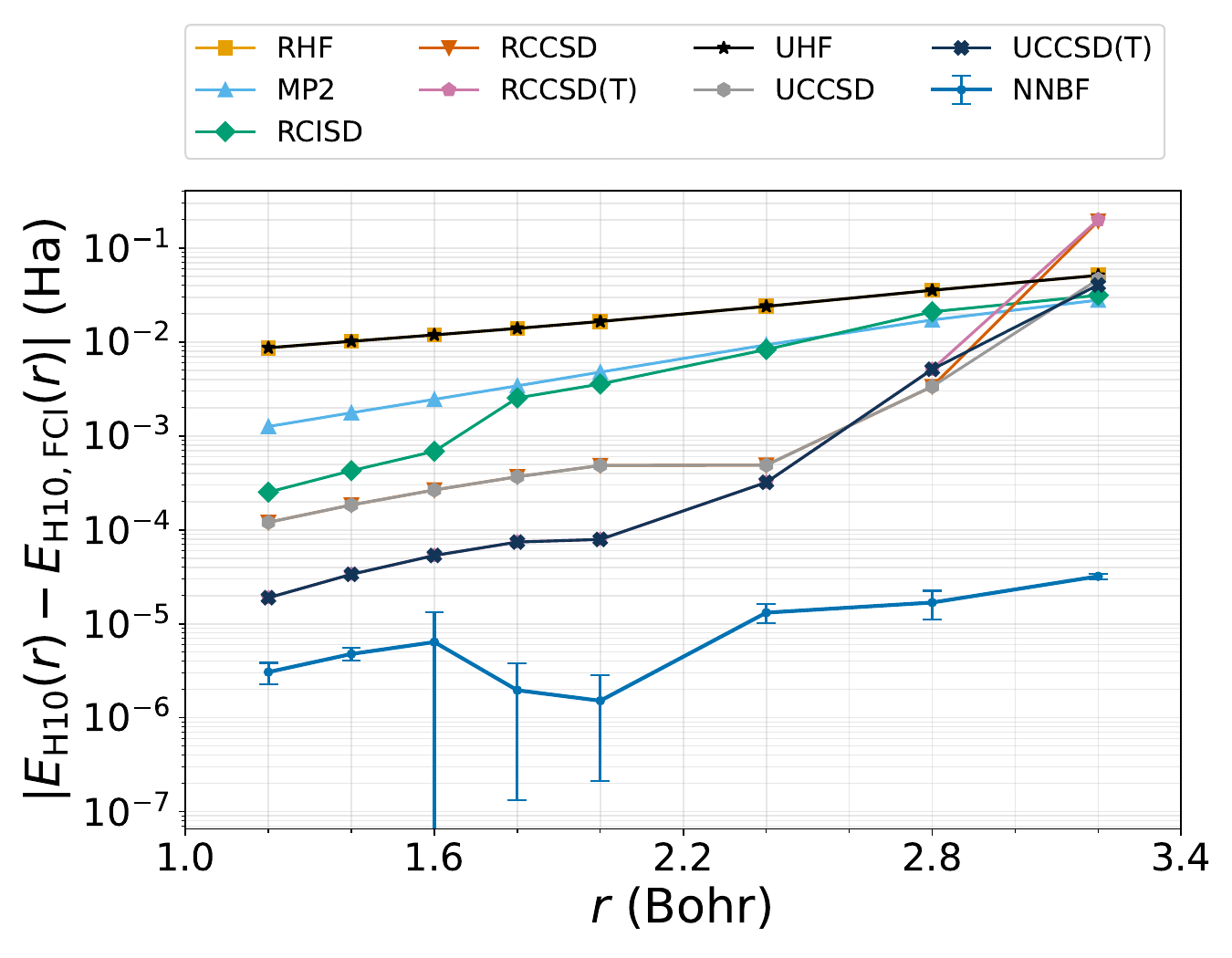}
    \caption{Potential energy curve for a H$_{10}$ under PBC using the STO-6G basis set. A unit cell of two hydrogen atoms is used, and the H$_{10}$ chain is modeled as a $5\times1\times1$ supercell. The energy errors of our NNBF ansatz and conventional quantum chemistry methods [HF, MP2, CISD, CCSD, CCSD(T)] are plotted relative to the exact FCI reference energy. The reported NNBF energies are obtained following the exact same training and evaluation protocol detailed in Fig.~\ref{fig:H10_OBC_curve}. Data for all other benchmark methods are computed using \textit{PySCF}~\cite{pyscf}.}
    \label{fig:H10_PBC_curve}
\end{figure}

We then transition to periodic boundary conditions (PBC) using a unit cell of two hydrogen atoms, presenting the STO-6G dissociation curve for an H$_{10}$ chain (modeled as a $5\times1\times1$ supercell). Figure~\ref{fig:H10_PBC_curve} displays a similar trend to the OBC case: NNBF maintains excellent agreement with FCI energies across the entire potential energy surface. It consistently outperforms standard coupled-cluster techniques, particularly in the bond-breaking regime where strong static correlation causes CCSD to fail.

Finally, we perform a TDL extrapolation for the periodic system near its equilibrium geometry. To account for finite-size scaling, the even and odd $N_k$ curves are fitted independently to the functional form $E = E_0 + B N^{-2}$. At every momentum mesh size $N_k$ (a unit cell with 2 hydrogen atoms is used), NNBF achieves a lower variational energy than all benchmarked conventional methods, including CCSD(T). Consequently, our approach converges to a lower extrapolated TDL energy, highlighting the superior accuracy and stability of the NNBF framework for periodic solids.

\begin{figure}[htbp!]
    \centering
    \includegraphics[width=\linewidth]{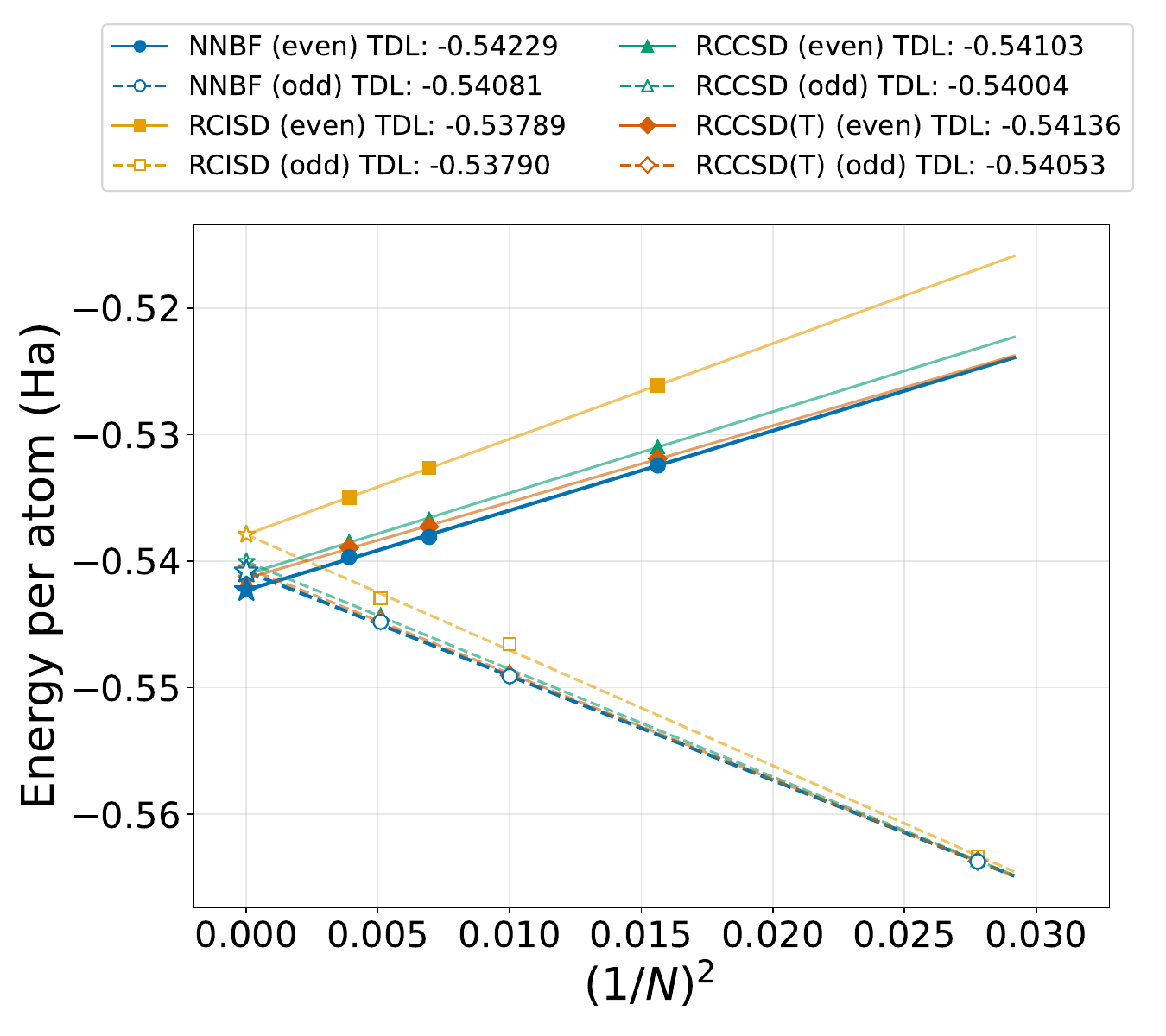}
    \caption{Ground-state energy convergence toward the thermodynamic limit (TDL) for the periodic hydrogen chain at an atomic separation of $r = 1.8$~Bohr in STO-6G basis set. Calculations are performed using our NNBF ansatz alongside conventional quantum chemistry methods [CISD, CCSD, CCSD(T)]. The system is modeled using a unit cell of two hydrogen atoms. The upper and lower branches for each method correspond to supercells containing an even and odd number of unit cells, respectively. The extrapolation to the macroscopic limit ($N \to \infty$) is carried out by independently fitting the even and odd sequences to the finite-size scaling formula $E(N) = E_0 + B N^{-2}$. The reported NNBF energies are obtained following the exact same training and evaluation protocol detailed in Fig.~\ref{fig:H10_OBC_curve}. Data for all other benchmark methods are computed using \textit{PySCF}~\cite{pyscf}.}
    \label{fig:PBC_TDL_r1.8}
\end{figure}

\subsection{Periodic 2D Graphene and 3D Silicon}

Next, we demonstrate the capability of our approach on ab-initio 2D and 3D periodic solids, specifically hexagonal graphene and face-centered cubic (FCC) silicon. Calculations for both materials employ the GTH-DZV basis set alongside GTH-Padé pseudopotentials, modeling a two-atom primitive unit cell. For graphene, we use a $2\times2\times1$ supercell, freezing the lowest 12 spatial orbitals to leave 16 active spatial orbitals. For silicon, we perform a pure $1\times1\times1$ supercell calculation utilizing the full orbital space. Consequently, both setups yield an identical active space size consisting of 8 electrons distributed across 32 spin-orbitals.

\begin{figure}[htbp!]
    \centering
    \includegraphics[width=\linewidth]{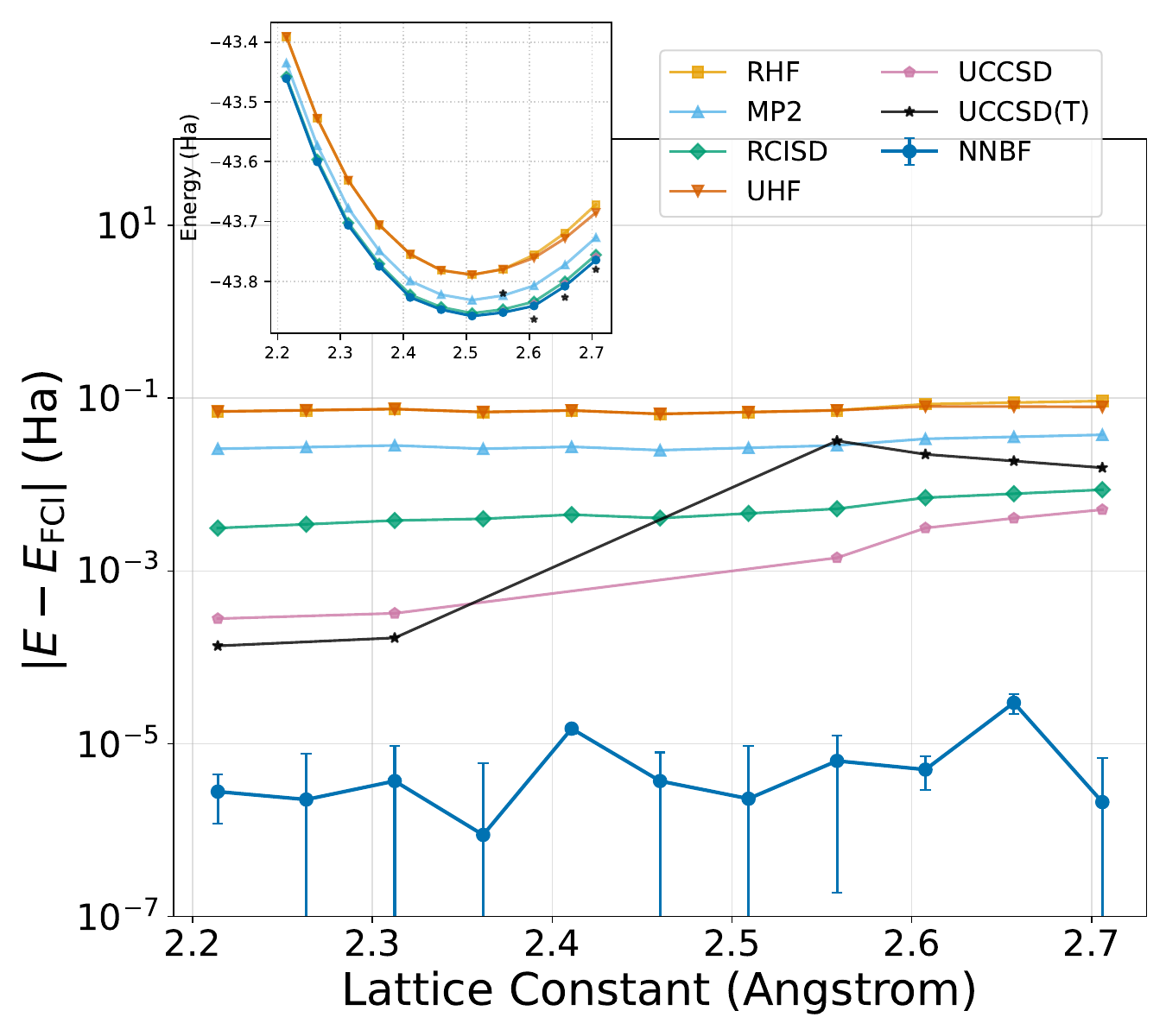}
    \caption{Ground-state potential energy curve for periodic 2D hexagonal graphene as a function of the lattice constant. Calculations are performed using the GTH-DZV basis set and GTH-Padé pseudopotentials with a $2\times2\times1$ supercell whose unit cell is a two-atom basis. The lowest 12 spatial orbitals are frozen, yielding an active space of 8 electrons across 32 spin-orbitals. The energy errors of our NNBF ansatz, constructed with canonical Hartree--Fock (HF) orbitals, and conventional quantum chemistry methods [HF, MP2, CISD, CCSD, CCSD(T)] are plotted relative to the exact FCI reference energy. Data points for certain coupled-cluster methods are omitted at specific lattice constants where the conventional calculations failed to converge.}
    \label{fig:graphene_curve}
\end{figure}

\begin{figure}[htbp!]
    \centering
    \includegraphics[width=\linewidth]{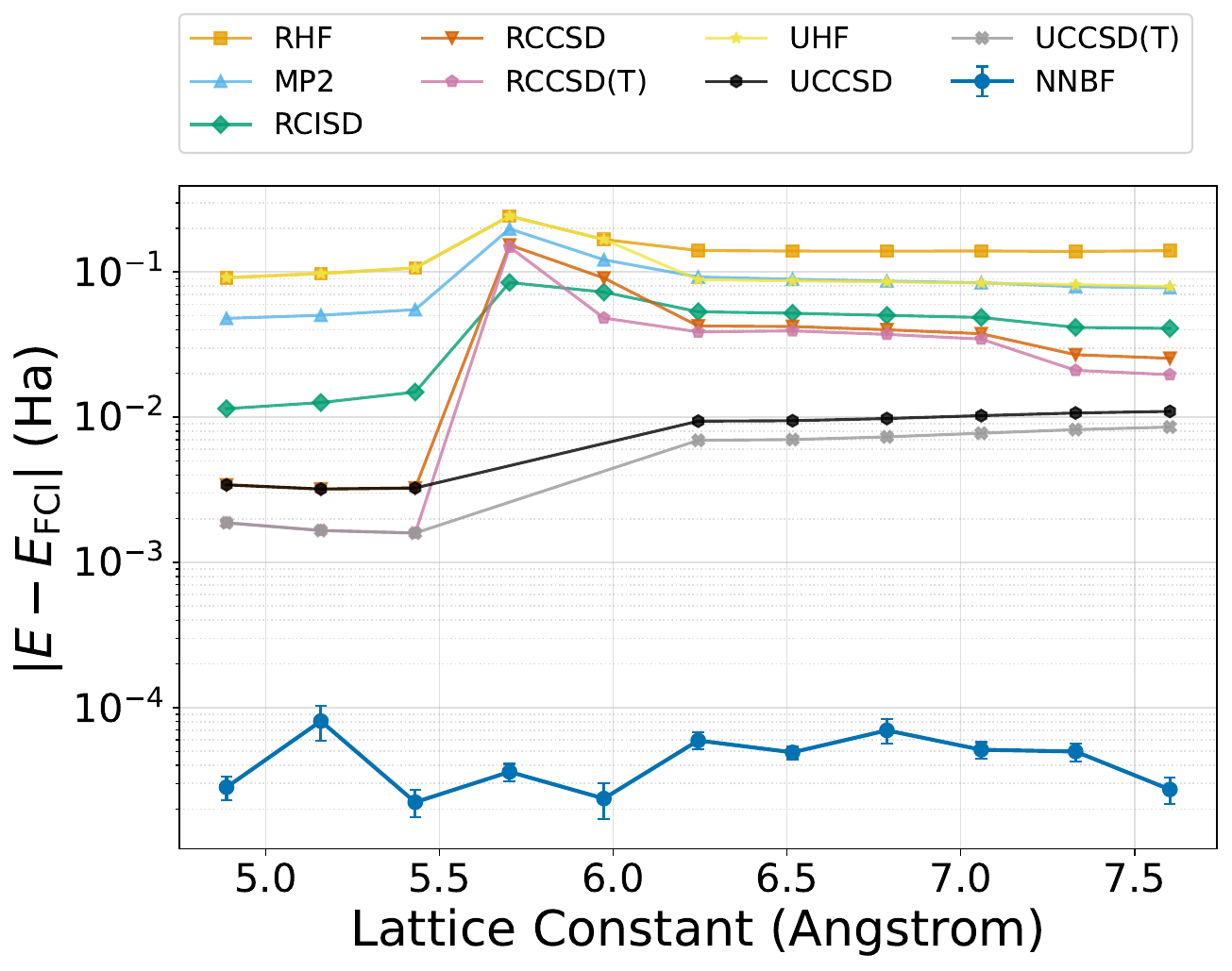}
    \caption{Ground-state potential energy curve for periodic 3D face-centered cubic (FCC) silicon as a function of the lattice constant. Calculations are performed using a $1\times1\times1$ supercell whose unit cell is a two-atom basis. GTH-DZV basis set and GTH-Padé pseudopotentials are used here. The full spatial orbital space is utilized without freezing any orbitals, resulting in an active space of 8 electrons across 32 spin-orbitals. The energy errors of our NNBF ansatz, constructed with canonical Hartree--Fock (HF) orbitals, and conventional quantum chemistry methods [HF, MP2, CISD, CCSD, CCSD(T)] are plotted relative to the exact FCI reference energy. Data points for certain coupled-cluster methods are omitted at specific lattice constants where the conventional calculations failed to converge.}
    \label{fig:silicon_curve}
\end{figure}

The potential energy curves—plotted as deviations from the exact FCI energy—are presented for graphene in Fig.~\ref{fig:graphene_curve} and for silicon in Fig.~\ref{fig:silicon_curve}. We benchmark our NNBF ansatz against conventional quantum chemistry techniques. The results demonstrate that NNBF maintains excellent agreement with the FCI reference energies across all sampled geometries. In stark contrast, gold-standard methods such as CCSD(T) perform poorly and even struggle to converge at certain atomic separations. This stark difference further highlights the superior accuracy and stability of the algorithmically enhanced NNBF framework for computing the electronic structure of periodic solids.

\subsection{Ablation study}\label{sec:ablation_study}

\subsubsection{Effectiveness of the Importance Proxy in Pruning}

This subsection demonstrates how the two-stage pruning strategy (introduced in Sec.~\ref{sec:algorithmic_improvement}) improves energy accuracy at almost no additional computational cost compared to the original intermittent target selection (ITS) method from Ref.~\onlinecite{Liu2025}.  The key advantage lies in providing a more representative pool space $\mathcal{P}$ from which to construct the target space $\mathcal{U}$.  To clearly compare the two approaches, we will perform an ablation study.   To facilitate a direct comparison, we first cast the original ITS algorithm into the language of our two-stage framework for a desired target space size $\abs{\mathcal{U}}$. Because the original ITS method does not utilize an importance-score proxy (omitting the first pruning stage entirely), it must operate with a core space of size $l\abs{\mathcal{U}}/N_{conn}$. This ensures its connected space $\mathcal{C}$—which acts as the pool space $\mathcal{P}$ in our terminology—contains $l\abs{\mathcal{U}}$ configurations. The exact NNBF amplitudes are then evaluated, and this space is trimmed by a factor of $l$ to yield the final target space $\mathcal{U}$.

With this formulation established, we perform an ablation study on an open boundary H$_{12}$ chain (STO-6G basis, $r=2.4$ Bohr). We fix the target space size at $\abs{\mathcal{U}} = 128 N_{conn}/l \approx 38,805$ and the sample size at $\abs{\mathcal{S}}=4096$, varying only the intermediate pool size $\abs{\mathcal{P}}$. To isolate the physical benefit of our heuristic from the simple mathematical benefit of deduplication, we introduce a ``random selection'' baseline where the importance score in Eq.~\eqref{eq:importance_score} is replaced with a random noise. Both our importance proxy and the random baseline utilize a fixed core size of $\abs{\mathcal{V}}=512$. In contrast, the original unpruned ITS method dynamically adjusts its core size ($\abs{\mathcal{V}} = \abs{\mathcal{P}}/N_{conn}$) to achieve the desired pool size.

\begin{figure}[htbp!]
    \centering
    \includegraphics[width=\linewidth]{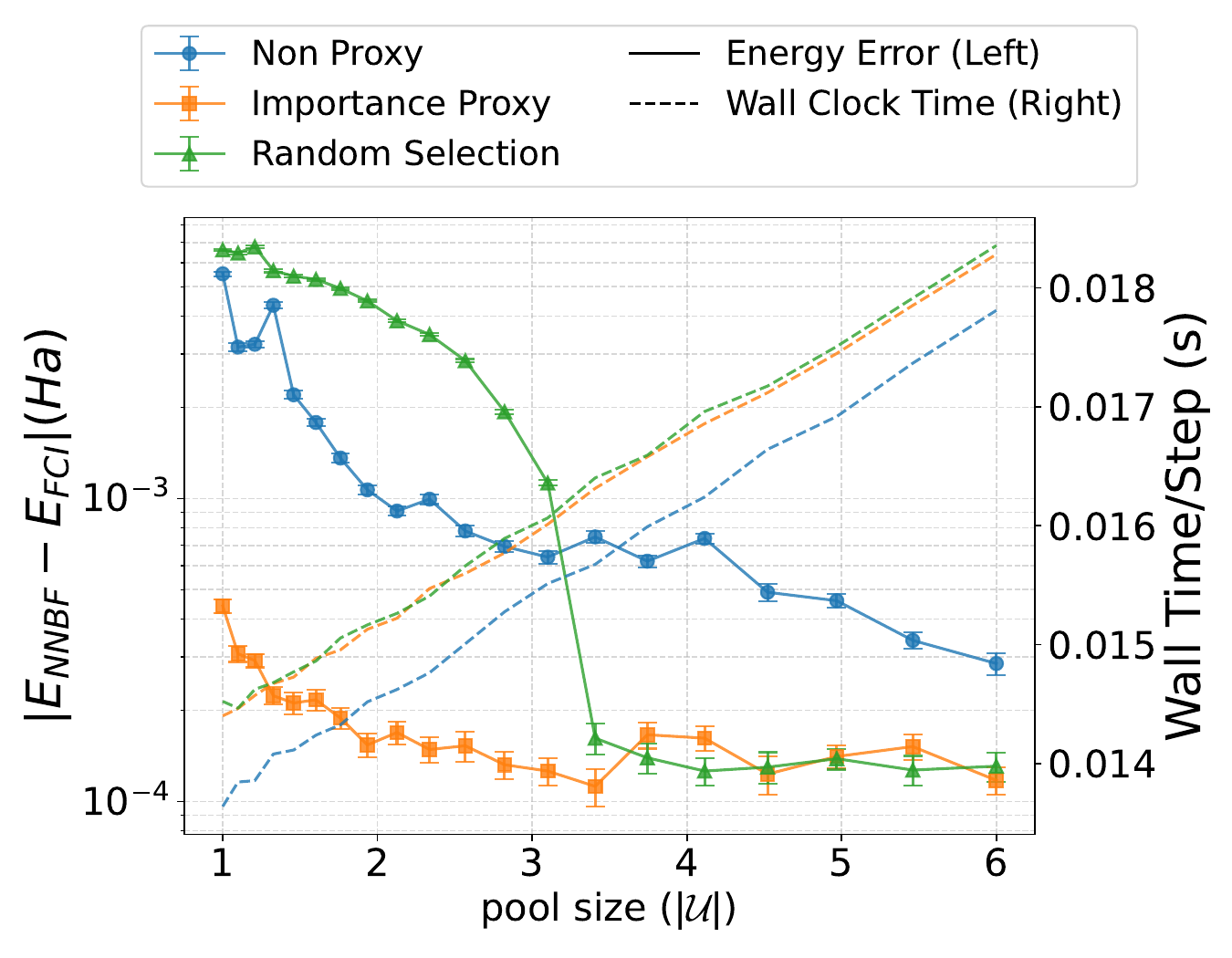}
    \caption{Ablation study evaluating the effect of the importance proxy on the energy accuracy of an open boundary H$_{12}$ using the STO-6G basis set and split localized PM molecular orbitals at an atomic separation of $r = 2.4$~Bohr. The energy error (and the average wall-clock time per optimization step) is plotted as a function of the intermediate pool space size ($\abs{\mathcal{P}}$), which is in the unit of target space size $\abs{\mathcal{U}}$ (fixed to 38805). The performance of our two-stage pruning strategy (using the physics-informed importance proxy) is compared against a random selection baseline and the vanilla unpruned intermittent target selection (ITS) method. Both the importance proxy and random expansion employ a fixed core size of $\abs{\mathcal{V}}=512$, whereas the unpruned method dynamically adjusts its core size to match the desired pool size.}
    \label{fig:pruning_ablation_study_curve}
\end{figure}

As demonstrated in Fig.~\ref{fig:pruning_ablation_study_curve}, our importance proxy yields profound improvements. In the leftmost points where the pool size equals the target space size ($\abs{\mathcal{P}} = \abs{\mathcal{U}}$)—meaning the second stage of exact NNBF amplitude pruning is bypassed entirely—our method achieves an order-of-magnitude improvement in energy accuracy over the vanilla unpruned method for a mere $5\%$ increase in runtime (0.0137 s to 0.0144 s). Furthermore, the energy rapidly converges when the pool size is just twice the target size. This incurs only a marginal additional runtime cost (from 0.0144 s to 0.0152 s), which accounts for both the expanded importance proxy calculations and the extra exact NNBF amplitude evaluations required prior to the second pruning stage. While the precise magnitude of these gains and runtime costs will naturally vary depending on the specific system, this striking ratio highlights the proxy's ability to efficiently capture the most physically relevant configurations from the connected space, allowing the second stage of exact NNBF amplitude pruning to select a highly representative target space $\mathcal{U}$.

The random expansion baseline exhibits poor performance—even worse than the vanilla unpruned method—until it surpasses a critical pool size, after which it performs excellently. We find that this critical pool size (approximately $3.5\abs{\mathcal{U}}$) corresponds precisely to the maximum capacity of the fully deduplicated connected space expanded from $\abs{\mathcal{V}}=512$. This behavior is expected: randomly deduplicating the connected space without accounting for graph connectivity or Hamiltonian matrix elements is an overly naive strategy. Importantly, this contrast confirms that the physics-informed metric in Eq.~\eqref{eq:importance_score} genuinely captures the underlying structure of the exact NNBF amplitudes, serving as a highly effective proxy.

\begin{figure}[htbp!]
    \centering
    \includegraphics[width=\linewidth]{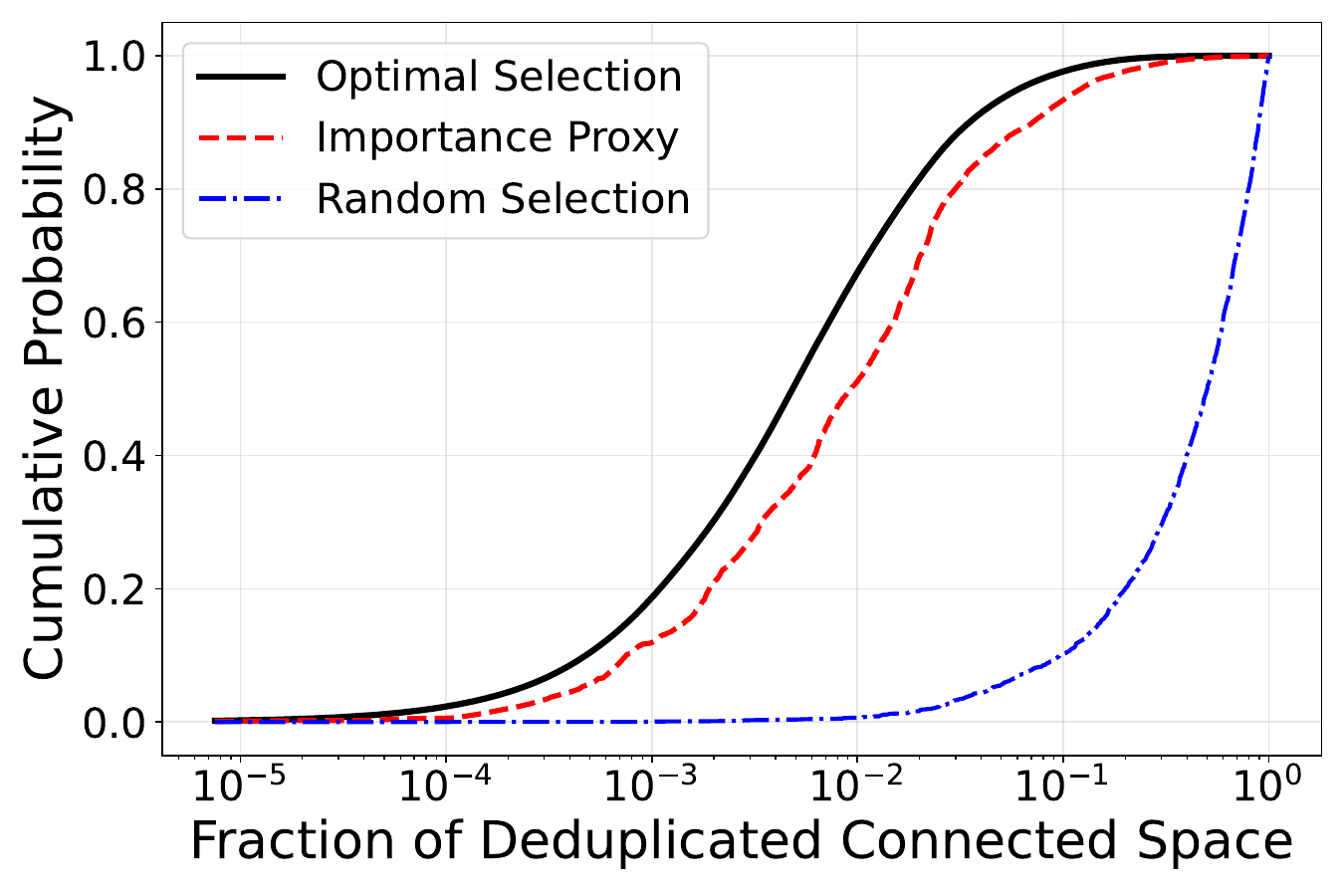}
    \caption{Cumulative probability mass of configurations selected from a fully deduplicated connected space ($\mathcal{C}_{unique}$) as a function of the selection fraction. This representative data point is taken from the H$_{12}$ chain ($r=2.4$~Bohr) analyzed in Fig.~\ref{fig:pruning_ablation_study_curve}. The optimal selection curve acts as the theoretical upper bound, generated by sorting configurations in descending order of their exact NNBF amplitude moduli. Building the intermediate pool space using our descending importance score proxy achieves a near-optimal selection curve. In contrast, the random selection baseline performs poorly, capturing a substantial probability mass only when it approaches the maximum capacity of $\mathcal{C}_{unique}$. This confirms that the physics-informed importance score successfully predicts the hierarchy of the exact NNBF amplitudes.}
    \label{fig:pruning_cmf}
\end{figure}

This conclusion is further validated by Fig.~\ref{fig:pruning_cmf}. Taking a representative data point from Fig.~\ref{fig:pruning_ablation_study_curve}, we generate a connected space $\mathcal{C}$ from the final core space, deduplicate it to form $\mathcal{C}_{unique}$, and compute the exact NNBF amplitudes for all its elements. The theoretical ``optimal selection'' for the pool space is formed by sorting these configurations in descending order of their exact amplitude moduli. Figure~\ref{fig:pruning_cmf} clearly demonstrates that cumulatively building the pool space via our descending importance score achieves a near-optimal selection curve. In contrast, the random selection only captures a non-trivial amount of probability mass when it approaches the full capacity of $\mathcal{C}_{unique}$—reiterating that physics-agnostic deduplication is inefficient. Ultimately, these results confirm that the proxy importance score successfully and cheaply predicts the hierarchy of the exact NNBF amplitudes.

\subsubsection{Effect of Basis Set}

Ref.~\onlinecite{Liu2025} demonstrated that employing coupled-cluster singles and doubles (CCSD) natural orbitals provides a marginal improvement in energy accuracy over canonical HF orbitals, as shown in a limited ablation study for Li$_2$O in the STO-3G basis set at a single geometry. Here, we present a more comprehensive ablation study examining the dissociation curve of an open boundary H$_{12}$ chain in the STO-6G basis set. We maintain a fixed core space size ($\abs{\mathcal{V}}=2048$), sample size ($\abs{\mathcal{S}}=4096$), training setup, and NNBF architecture, varying only the choice of molecular orbitals: canonical Hartree–Fock (HF), CCSD natural orbitals (NO), and split localized Pipek–Mezey (PM) orbitals.

\begin{figure}[htbp!]
    \centering
    \includegraphics[width=\linewidth]{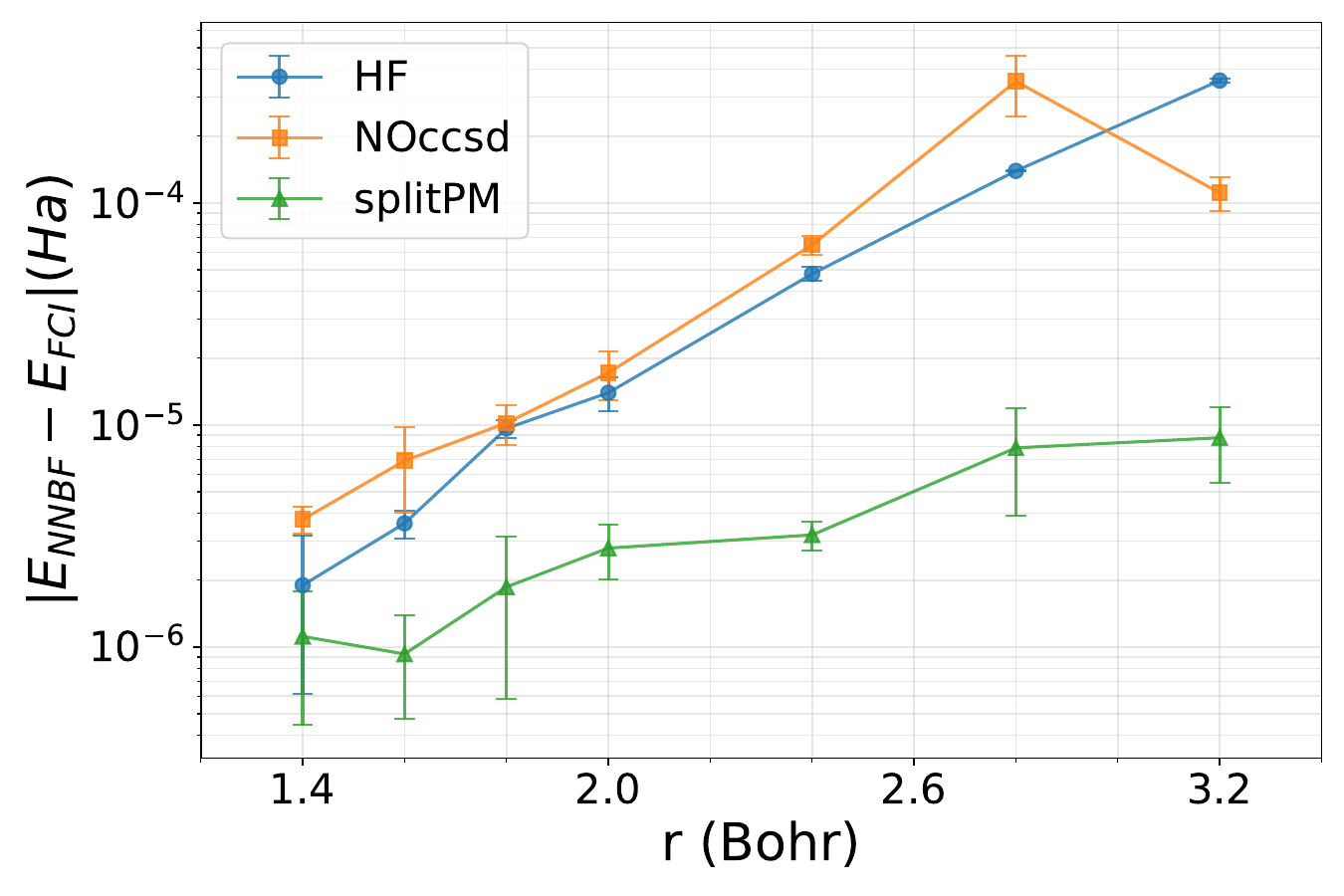}
    \caption{Ablation study evaluating the effect of the single-particle basis on the potential energy curve of a H$_{12}$ under OBC using the STO-6G basis set. The energy errors of the NNBF ansatz, plotted relative to the exact FCI reference, are compared across three different choices of molecular orbitals: canonical Hartree--Fock (HF), CCSD natural orbitals (NO), and split localized Pipek--Mezey (PM) orbitals. To isolate the effect of the orbitals, all three models employ an identical training setup with a fixed core space size of $\abs{\mathcal{V}}=2048$ and sample size of $\abs{\mathcal{S}}=4096$. While all orbital choices perform excellently at shorter interatomic distances where dynamic correlation dominates, the canonical HF and CCSD natural orbitals degrade significantly in the strongly correlated stretched regime. In contrast, the split localized PM orbitals maintain robust, high accuracy across the entire dissociation curve.}
    \label{fig:basis_set_ablation_study}
\end{figure}

Figure~\ref{fig:basis_set_ablation_study} demonstrates that at shorter interatomic distances, where dynamic correlation dominates, our method performs exceptionally well across all three orbital bases. However, as the interatomic distance increases and static correlation becomes dominant, the performance using canonical HF and CCSD natural orbitals begins to degrade. This behavior is expected, as the underlying mean-field and single-reference methods themselves severely struggle in this strongly correlated regime. In contrast, while the model using split PM orbitals also exhibits a slight decrease in accuracy at stretched geometries, it remains highly accurate overall. This robustness is directly attributed to the localized nature of the PM method, with the split-localization strategy successfully preserving the physical advantages of both canonical HF and localized orbitals. 

The purpose of this ablation study is not to definitively declare one molecular orbital basis as universally superior. Rather, it is to highlight that the choice of the single-particle basis plays a highly significant, yet often overlooked, role in the training performance and ultimate accuracy of neural quantum states.

\section{Conclusion}

In this work, we have successfully extended the scalable NNBF framework from ref.~\onlinecite{Liu2025} to real periodic solid-state materials by introducing an efficient two-stage pruning algorithm based on a physics-informed importance score. On both open and periodic 1D hydrogen chains, our method matches or exceeds the performance of established state-of-the-art ab-initio methods such as AFQMC and DMRG. Crucially, in strongly correlated bond-breaking regimes where gold-standard coupled-cluster methods like CCSD and CCSD(T) break down, the NNBF ansatz remains highly robust, enabling smooth and accurate extrapolations to the thermodynamic limit. We further demonstrated the method's capability to scale to real extended materials by successfully computing the electronic structure of 2D hexagonal graphene and 3D face-centered cubic silicon. 

Through detailed ablation studies, we validated the efficacy of our two-stage pruning strategy over naive deduplication methods. By utilizing a computationally cheap importance proxy to construct a highly representative intermediate subspace, we ensure that the subsequent, exact NNBF amplitude evaluations are concentrated only on the most physically relevant configurations, drastically improving energy accuracy with negligible runtime cost. Furthermore, we highlighted the profound impact that the choice of single-particle basis sets exerts on NQS training, showing that preserving orbital localization (such as via split Pipek--Mezey orbitals) is vital for maintaining accuracy as systems transition into regimes dominated by strong static correlation.

Looking forward, this framework lays the foundation for promising research directions. Given the sensitivity of the training performance to the single-particle basis, integrating orbital optimization—such as iterative orbital rotation\cite{Tubman2020} or directly parameterizing orbital rotation\cite{Javier2023}—into the NQS framework warrants thorough investigation. It would also be interesting to explore the use of other architectures for periodic solid ab-initio NNBF.
\cite{Lv2025,Shang2025,Di2025}. We anticipate that the techniques developed in this study will enable more efficient and reliable NQS optimization, expanding the range of practical applications for computing the electronic structure of complex periodic solids.


\begin{acknowledgments}
This work made use of the Illinois Campus Cluster, a computing resource that is operated by the Illinois Campus Cluster Program (ICCP) in conjunction with the National Center for Supercomputing Applications (NCSA) and which is supported by funds from the University of Illinois Urbana-Champaign. We also acknowledge support from the NSF Quantum Leap Challenge Institute for Hybrid Quantum Architectures and Networks (NSF Award 2016136).
\end{acknowledgments}

\appendix
\section{Experimental Setup}\label{appx:experimental_setup}

The core architecture of our NNBF ansatz relies on a multi-layer perceptron (MLP) parameterized by $h$ hidden units and $L$ hidden layers, which ultimately outputs $D$ configuration-dependent determinants. To facilitate stable gradient flow in deeper networks ($L>1$), residual connections are explicitly included. The variational energy is minimized to approximate the true ground-state energy using the AdamW\cite{AdamW} optimizer, with all standard optimization hyperparameters detailed in Table~\ref{tab:hyperparameters_and_notations}. All reference electronic structure computations—including HF, CCSD, CCSD(T), and exact diagonalization (FCI)—are executed using the \textit{PySCF} library \cite{pyscf}.

To evaluate the final variational energy and its associated statistical uncertainty after the optimization phase is complete, we perform an independent Markov Chain Monte Carlo (MCMC) inference run. We deploy $N_w=1024$ independent walkers to sample the unnormalized probability density, defined as $\Bar{p}_\theta(\bm{x})=|\psi_{\theta, NNBF}(\bm{x})|^2$, via the Metropolis–Hastings algorithm. The proposal mechanism generates configurations by executing spin-conserving single-particle excitations (i.e., swapping an occupied spin-orbital with an unoccupied orbital of identical spin). To effectively mitigate autocorrelation between consecutive states, the Markov chains are thinned, with samples recorded only after an interval of $K_1=10N_e$ steps.

To prevent the walkers from becoming trapped in local minima and to guarantee a thorough exploration of the phase space, the MCMC initialization adopts an ensemble-based approach inspired by Ref.~\onlinecite{ForemanMackey2013}. Walkers are initially seeded across a distribution constructed from the eight most heavily weighted configurations identified during the final optimization step. The walkers are then thermalized through a strict burn-in phase lasting $K_2=100K_1$ steps. 

Once equilibrium is achieved across the ensemble, we harvest $T=1000$ uncorrelated configurations per walker. The ultimate energy expectation value is calculated as the global average across all $T \times N_w$ collected samples. The statistical error is subsequently reported as the standard error of this global mean, evaluated as $\sqrt{\text{Var}(E)/(T \times N_w)}$. All final NNBF energies presented throughout this study are derived from this rigorous post-training MCMC evaluation.

\begin{table*}[htbp!]
\centering
\begin{tabular}{lll}
\toprule
\textbf{Parameter / Symbol} & \textbf{Description} & \textbf{Default Value / Definition} \\
\midrule
\multicolumn{3}{l}{\textit{\textbf{A. Physical System \& Configuration Spaces}}} \\
\midrule
$N_e$ & Number of electrons & Problem-dependent \\
$N_o$ & Number of spin-orbitals & Problem-dependent \\
$\mathcal{H}$ & The physical Hilbert space & Problem-dependent \\
$\mathcal{V}$ & Core space & See Section \ref{sec:background} \\
$\mathcal{C}$ & Connected space & Generated from core space \\
$\mathcal{P}$ & Pool space & See Section \ref{sec:algorithmic_improvement} \\
$\mathcal{U}$ & Target space & See Section \ref{sec:background} \\
$\mathcal{S}$ & \makecell[l]{Sample set for energy estimations} & See Section \ref{sec:background} \\
\midrule
\multicolumn{3}{l}{\textit{\textbf{B. Model Architecture (MLP)}}} \\
\midrule
$D$ & Number of backflow determinants & 1 \\
$L$ & Number of hidden layers & 2 \\
$h$ & Number of hidden units per layer & 512\\
\midrule
\multicolumn{3}{l}{\textit{\textbf{C. Optimizer \& Training Schedule}}} \\
\midrule
Optimizer & Algorithm for parameter updates & AdamW \\
$\beta_1, \beta_2$ & AdamW exponential decay rates & 0.9, 0.999 \\
$\epsilon$ & AdamW epsilon for numerical stability & $1 \times 10^{-8}$ \\
$\lambda$ & AdamW weight decay & $1 \times 10^{-4}$ \\
Learning Rate ($t$) & Initial rate with decay over iteration $t$ & $10^{-3} \times (1+10^{-5}t)^{-1}$ \\
\midrule
\multicolumn{3}{l}{\textit{\textbf{D. Algorithmic Enhancements}}} \\
\midrule
$l$ & Speedup factor & $N_o - N_e/2$, see Section \ref{sec:background} \\
$r$ & Reduction factor & 4, see Section \ref{sec:algorithmic_improvement} \\
\midrule
\multicolumn{3}{l}{\textit{\textbf{E. Implementation Details}}} \\
\midrule
Framework & Core computational library & JAX \\
Precision & Floating-point precision & float32 \\
Energy Unit & Standard unit for energy values & Hartree \\
\midrule
\multicolumn{3}{l}{\textit{\textbf{F. Post-Training MCMC Inference}}} \\
\midrule
$M_{\text{init}}$ & Dominant configurations for initialization & 8 \\
$N_w$ (inference) & Number of MCMC walkers & 1024 \\
$K_2$ & Burn-in steps per walker & $100K_1$ \\
$K_1$ & Downsample interval (iterations) & $10N_e$ \\
$M$ & Samples collected per walker & 1000 \\
\bottomrule
\end{tabular}
\caption{Consolidated hyperparameters and notations used for all experiments, unless explicitly stated otherwise.}
\label{tab:hyperparameters_and_notations}
\end{table*}

\section{Thermodynamic Limit Extrapolation for Open Boundary Chains}\label{appx:OBC_TDL_extrapolation}

Here, we detail the procedure used to extrapolate the NNBF ground-state energies to the thermodynamic limit (TDL) for the hydrogen chain under open boundary conditions (OBC) using the STO-6G basis set. 

For each interatomic distance $r$, we execute three independent training runs using the NNBF ansatz constructed with split localized Pipek--Mezey (PM) molecular orbitals. Following the optimization phase, a preliminary post-training MCMC inference is performed to evaluate the variational energy of each trial. To avoid biased estimation, the model yielding the lowest energy is selected, and a completely independent, final MCMC inference is conducted on this optimal model to obtain the true, unbiased energy and its associated statistical uncertainty.

For a given distance $r$, the unbiased energies obtained from chains of varying lengths ($N$ atoms) are normalized by the number of atoms. We then perform a weighted least squares (WLS) fit of these per-atom energies against the inverse system size ($1/N$) using the finite-size scaling formula $E(N) = E_0 + A N^{-1} + B N^{-2}$. The macroscopic TDL energy is obtained by evaluating the limit as $N \to \infty$ (i.e., $1/N \to 0$), which corresponds to extracting the intercept $E_0$ alongside its propagated fitting uncertainty.

These extrapolated $E_0$ values are subsequently used to construct the TDL potential energy curve for the NNBF ansatz used in Figure \ref{fig:OBC_TDL_curve}. Data for all other benchmark methods, including conventional quantum chemistry methods [CCSD, CCSD(T)], AFQMC, and DMRG, are taken directly from Ref.~\onlinecite{Motta2017}.

\begin{figure*}[htbp]
    \centering
    \includegraphics[width=0.48\textwidth]{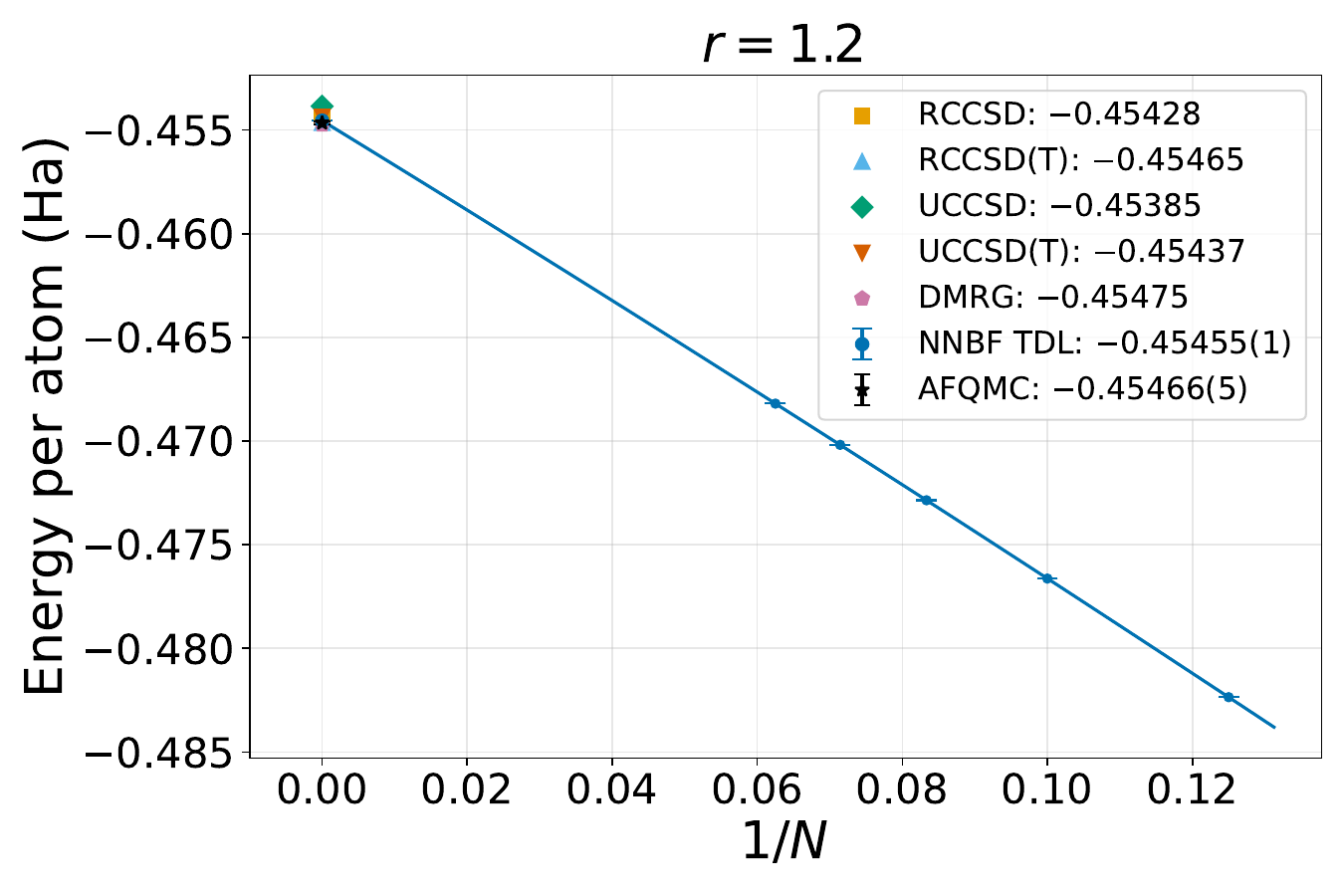}\hfill
    \includegraphics[width=0.48\textwidth]{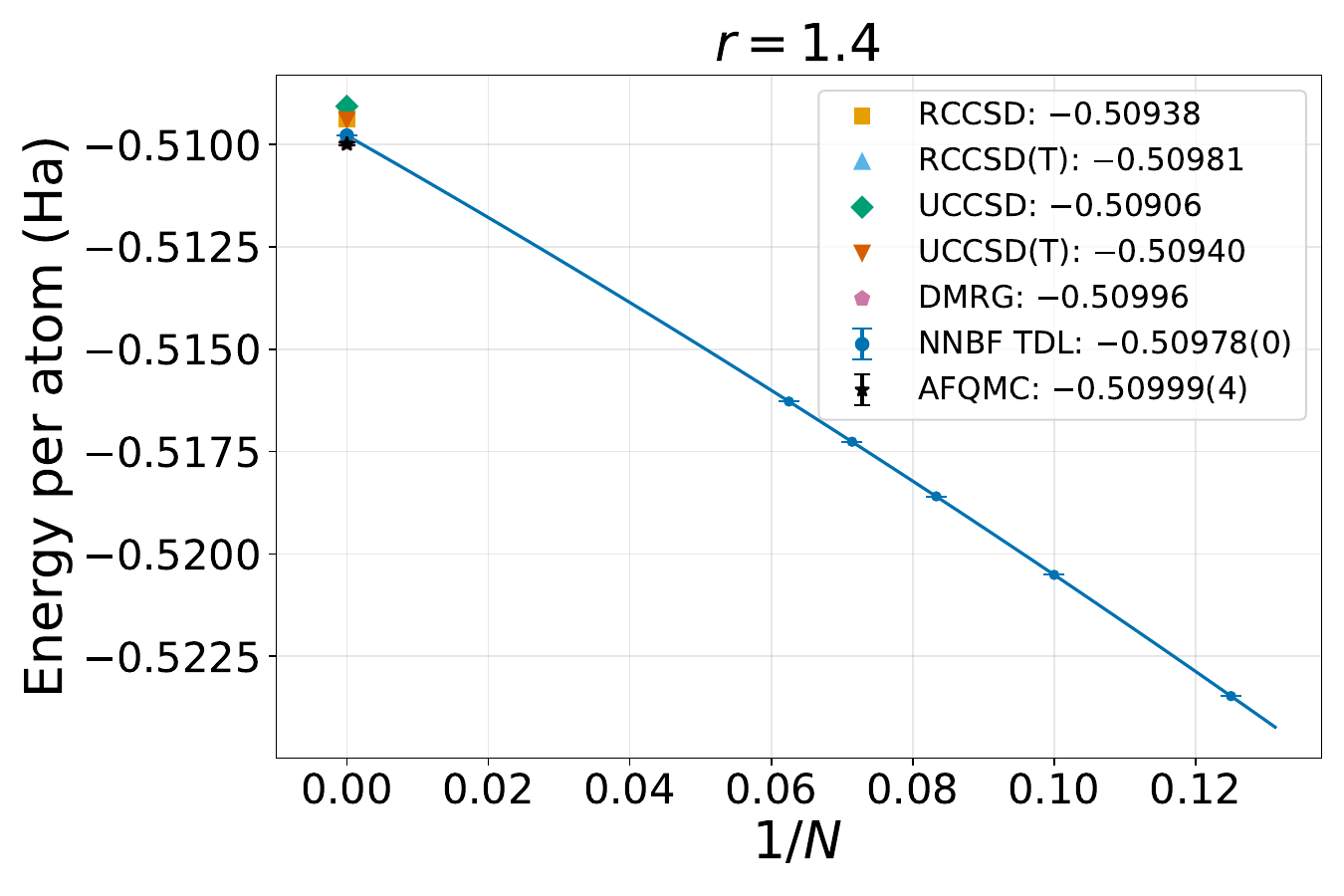}\\[\medskipamount]
    
    \includegraphics[width=0.48\textwidth]{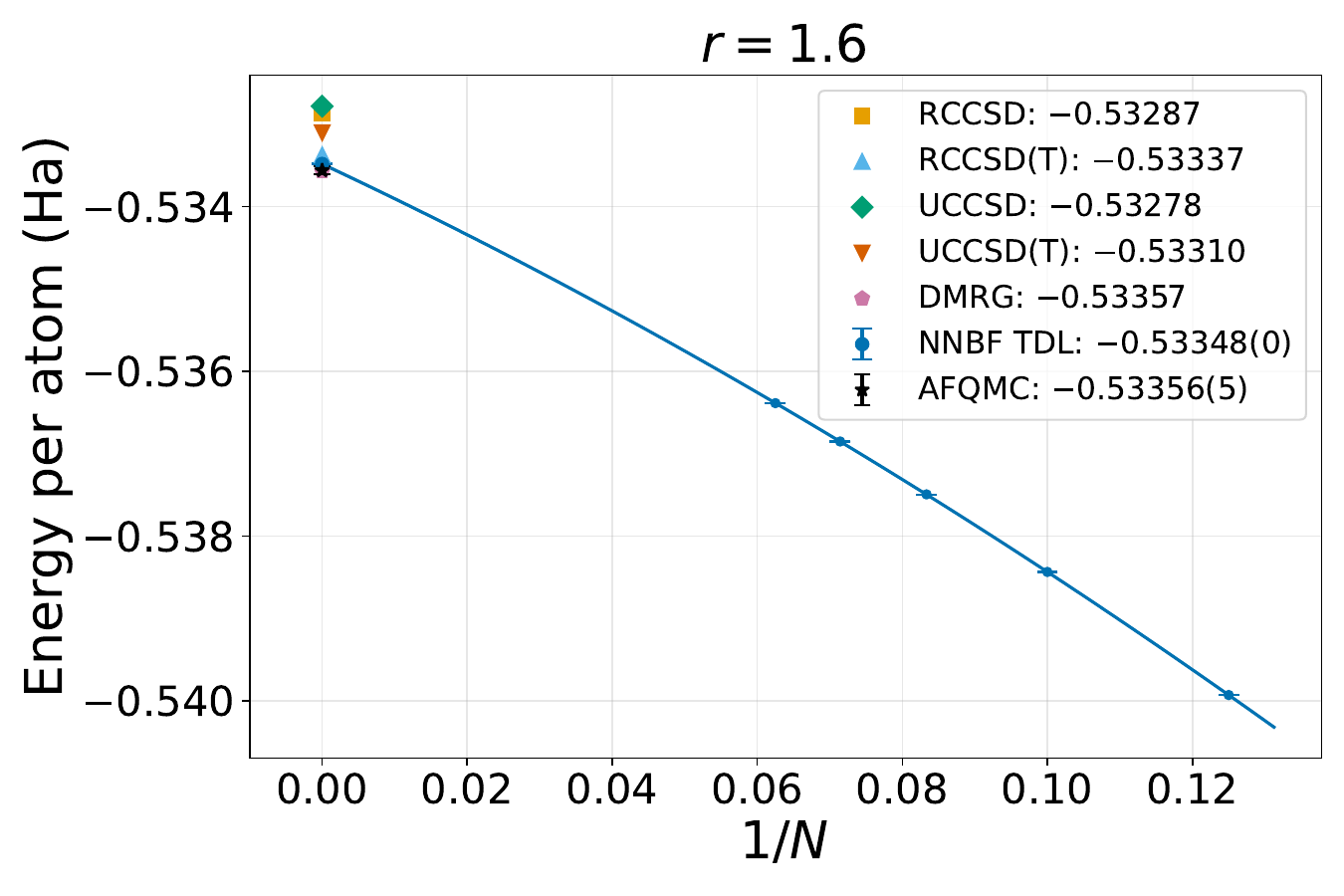}\hfill
    \includegraphics[width=0.48\textwidth]{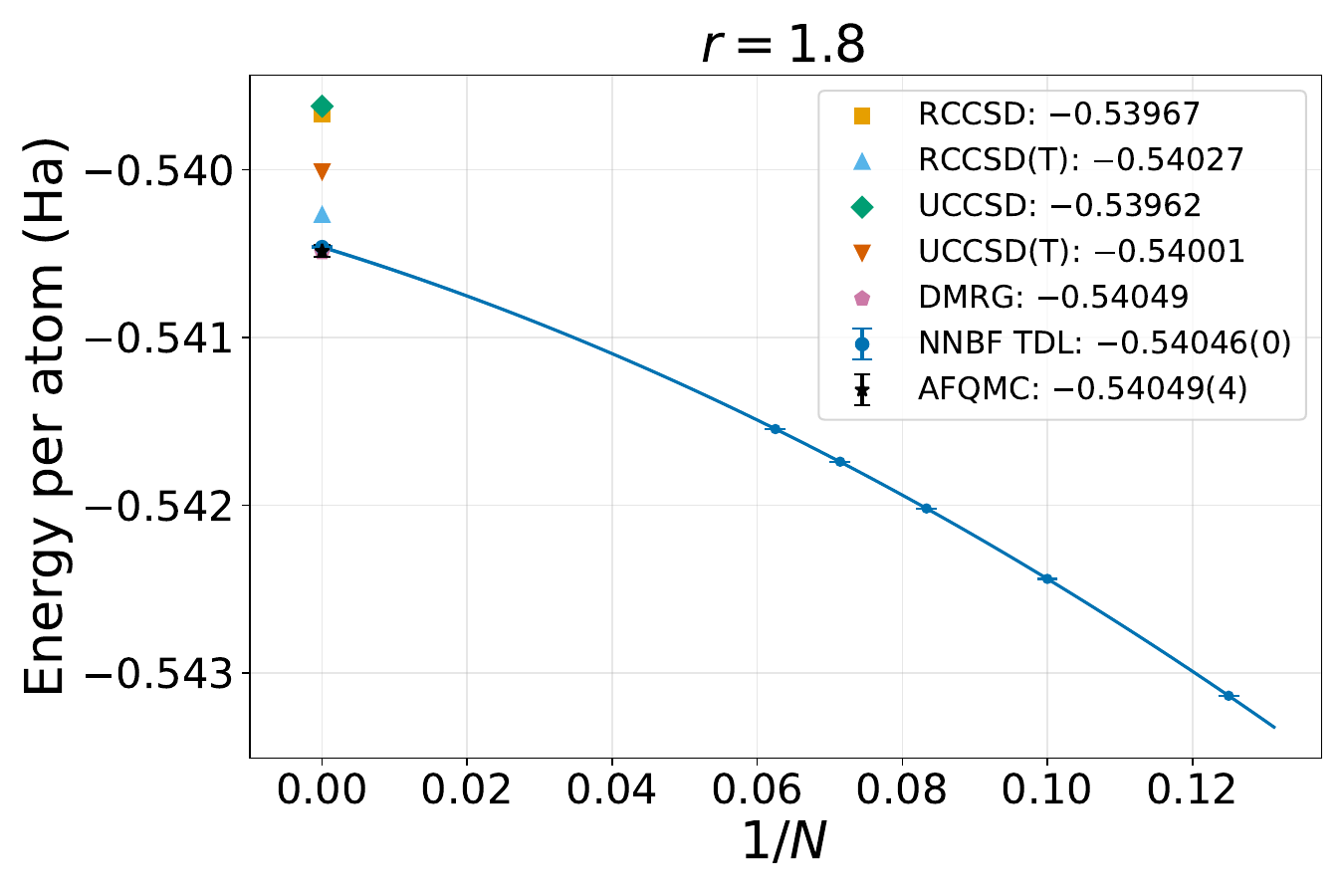}\\[\medskipamount]
    
    \includegraphics[width=0.48\textwidth]{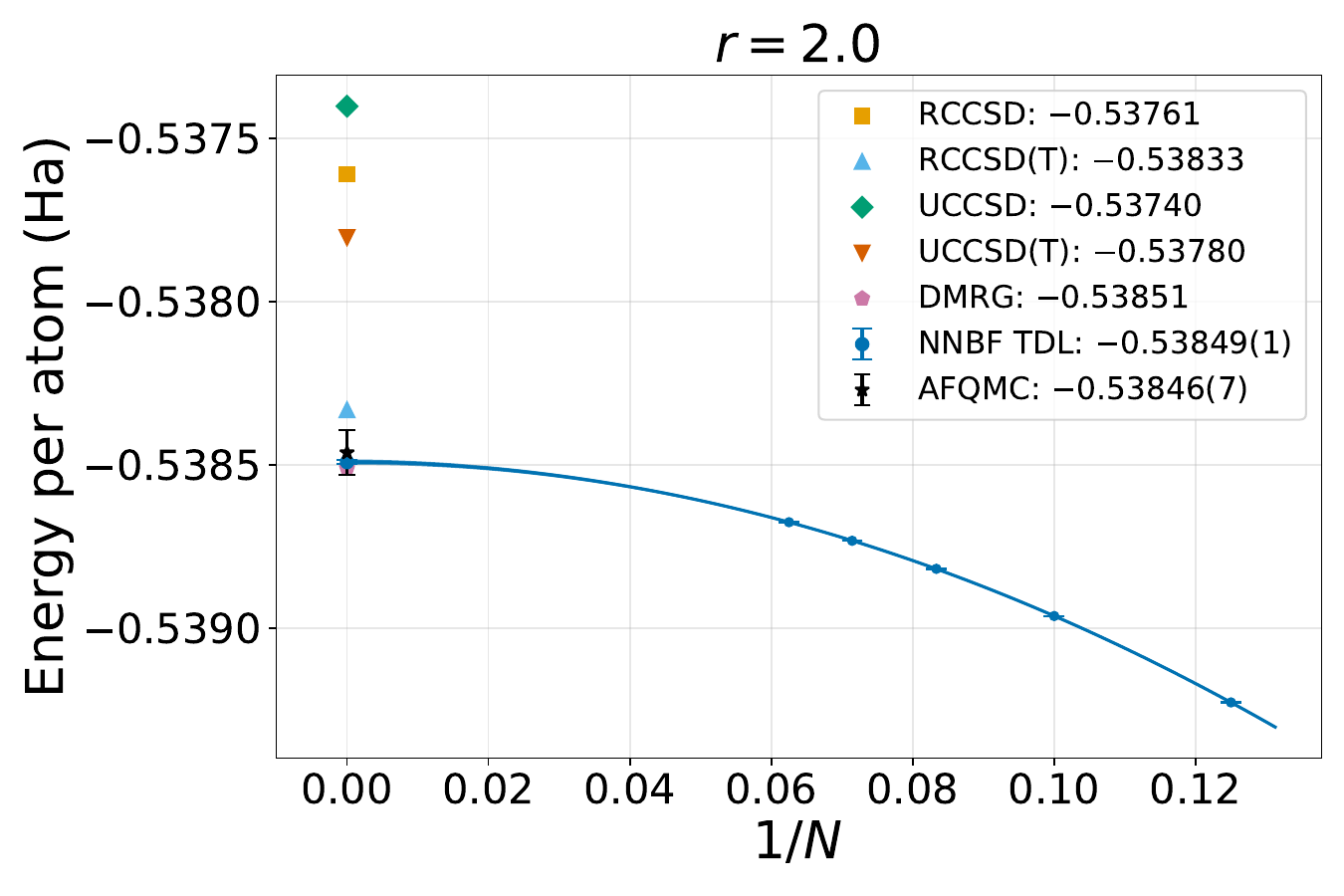}\hfill
    \includegraphics[width=0.48\textwidth]{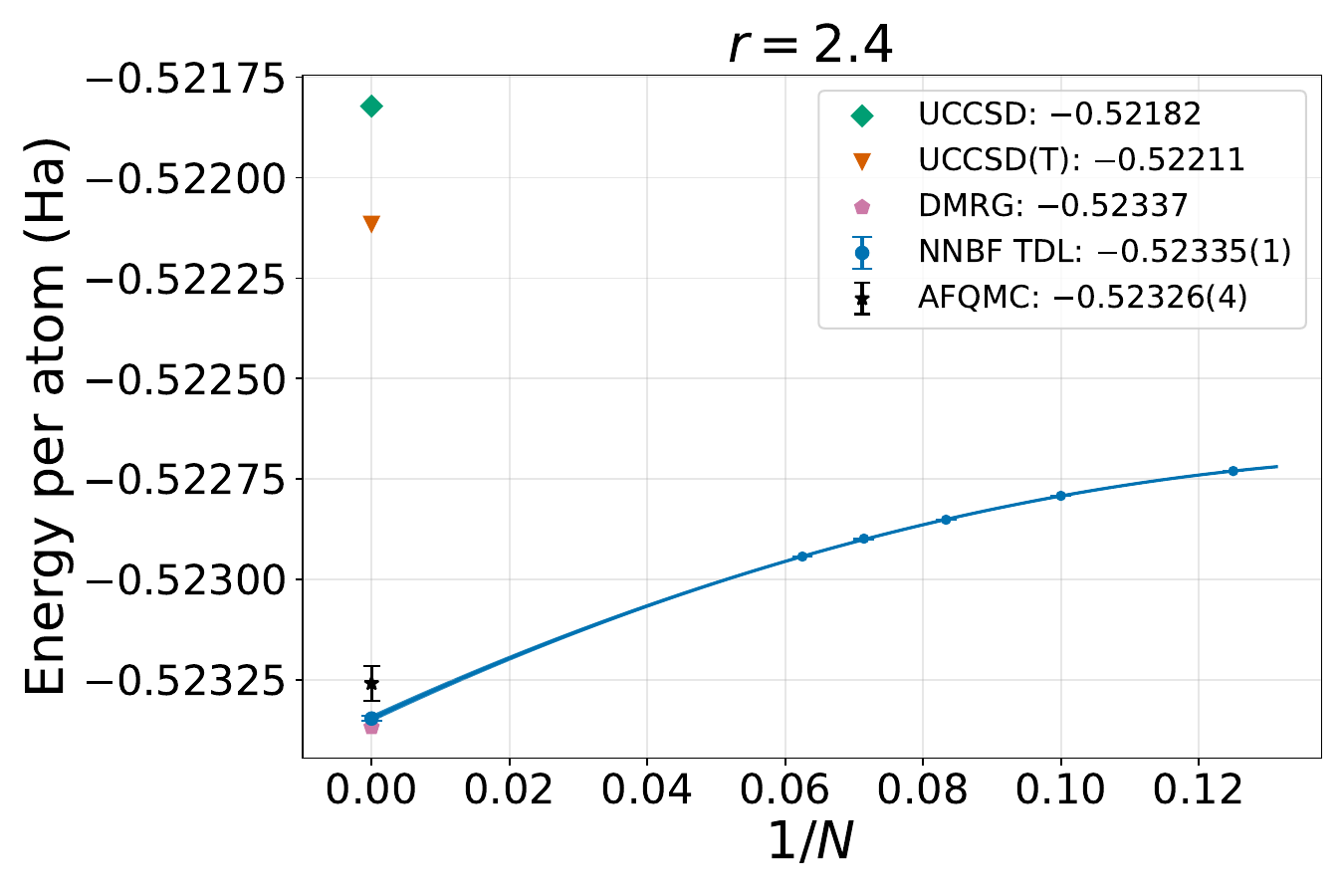}\\[\medskipamount]
    
    \includegraphics[width=0.48\textwidth]{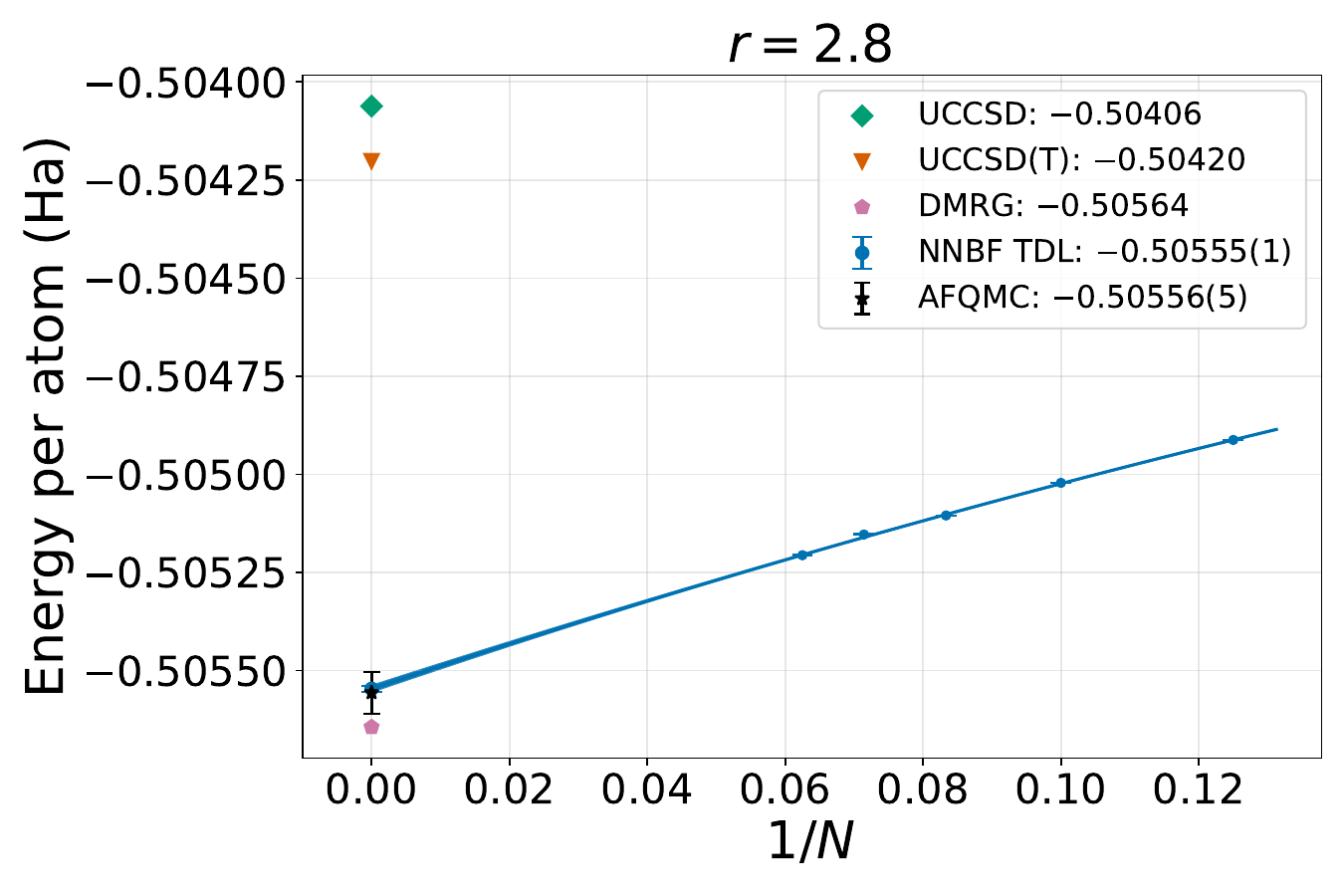}\hfill
    \includegraphics[width=0.48\textwidth]{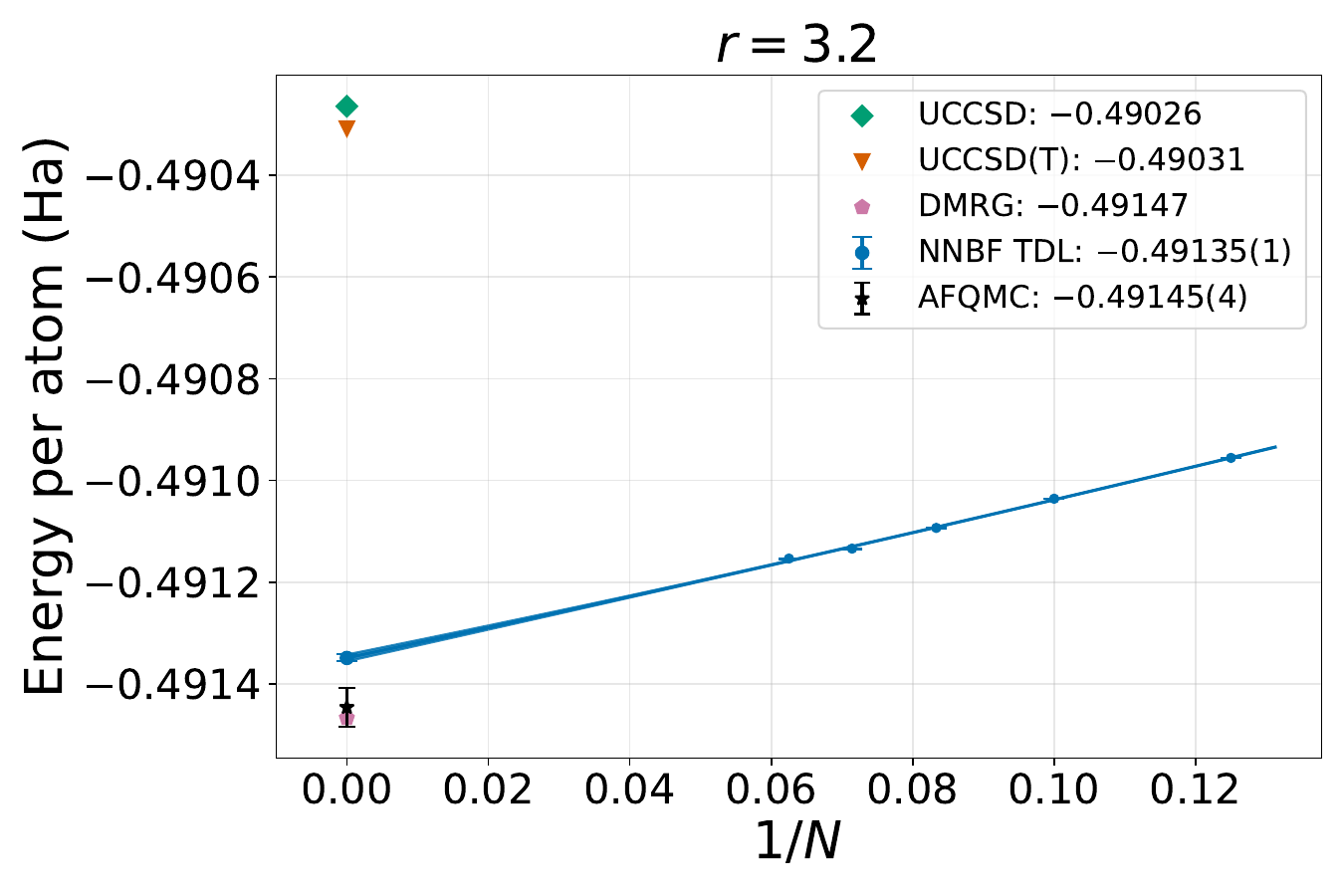}
    
    \caption{Finite-size scaling extrapolations to the thermodynamic limit (TDL) for the open boundary hydrogen chain at eight distinct atomic separations using the STO-6G basis set. The NNBF energy per particle is fitted to a second-order polynomial, and data for the other benchmark methods are taken from Ref.~\onlinecite{Motta2017}.}
    \label{fig:main_label}
\end{figure*}


\clearpage 
\bibliography{reference}

@article{RobledoMoreno2022,
  title = {Fermionic wave functions from neural-network constrained hidden states},
  volume = {119},
  ISSN = {1091-6490},
  url = {http://dx.doi.org/10.1073/pnas.2122059119},
  DOI = {10.1073/pnas.2122059119},
  number = {32},
  journal = {Proceedings of the National Academy of Sciences},
  publisher = {Proceedings of the National Academy of Sciences},
  author = {Robledo Moreno,  Javier and Carleo,  Giuseppe and Georges,  Antoine and Stokes,  James},
  year = {2022},
  month = aug 
}

@article{Zhang2003,
  title = {Quantum Monte Carlo Method using Phase-Free Random Walks with Slater Determinants},
  volume = {90},
  ISSN = {1079-7114},
  url = {http://dx.doi.org/10.1103/PhysRevLett.90.136401},
  DOI = {10.1103/physrevlett.90.136401},
  number = {13},
  journal = {Physical Review Letters},
  publisher = {American Physical Society (APS)},
  author = {Zhang,  Shiwei and Krakauer,  Henry},
  year = {2003},
  month = apr 
}

@article{Zhang1997,
  title = {Constrained path Monte Carlo method for fermion ground states},
  volume = {55},
  ISSN = {1095-3795},
  url = {http://dx.doi.org/10.1103/PhysRevB.55.7464},
  DOI = {10.1103/physrevb.55.7464},
  number = {12},
  journal = {Physical Review B},
  publisher = {American Physical Society (APS)},
  author = {Zhang,  Shiwei and Carlson,  J. and Gubernatis,  J. E.},
  year = {1997},
  month = mar,
  pages = {7464–7477}
}

@article{Blankenbecler1981,
  title = {Monte Carlo calculations of coupled boson-fermion systems. I},
  volume = {24},
  ISSN = {0556-2821},
  url = {http://dx.doi.org/10.1103/PhysRevD.24.2278},
  DOI = {10.1103/physrevd.24.2278},
  number = {8},
  journal = {Physical Review D},
  publisher = {American Physical Society (APS)},
  author = {Blankenbecler,  R. and Scalapino,  D. J. and Sugar,  R. L.},
  year = {1981},
  month = oct,
  pages = {2278–2286}
}

@article{Chan2016,
  title = {Matrix product operators,  matrix product states,  and ab initio density matrix renormalization group algorithms},
  volume = {145},
  ISSN = {1089-7690},
  url = {http://dx.doi.org/10.1063/1.4955108},
  DOI = {10.1063/1.4955108},
  number = {1},
  journal = {The Journal of Chemical Physics},
  publisher = {AIP Publishing},
  author = {Chan,  Garnet Kin-Lic and Keselman,  Anna and Nakatani,  Naoki and Li,  Zhendong and White,  Steven R.},
  year = {2016},
  month = jul 
}

@article{Yoshioka2021,
  title = {Solving quasiparticle band spectra of real solids using neural-network quantum states},
  volume = {4},
  ISSN = {2399-3650},
  url = {http://dx.doi.org/10.1038/s42005-021-00609-0},
  DOI = {10.1038/s42005-021-00609-0},
  number = {1},
  journal = {Communications Physics},
  publisher = {Springer Science and Business Media LLC},
  author = {Yoshioka,  Nobuyuki and Mizukami,  Wataru and Nori,  Franco},
  year = {2021},
  month = may 
}

@article{Shang2024-Solid,
  title = {Transformer-Based Neural-Network Quantum State Method for Electronic Band Structures of Real Solids},
  volume = {20},
  ISSN = {1549-9626},
  url = {http://dx.doi.org/10.1021/acs.jctc.4c00567},
  DOI = {10.1021/acs.jctc.4c00567},
  number = {14},
  journal = {Journal of Chemical Theory and Computation},
  publisher = {American Chemical Society (ACS)},
  author = {Fu,  Lizhong and Wu,  Yangjun and Shang,  Honghui and Yang,  Jinlong},
  year = {2024},
  month = jul,
  pages = {6218–6226}
}

@article{Shang2024-DMET,
  title = {Quantum embedding method with transformer neural network quantum states for strongly correlated materials},
  volume = {10},
  ISSN = {2057-3960},
  url = {http://dx.doi.org/10.1038/s41524-024-01406-3},
  DOI = {10.1038/s41524-024-01406-3},
  number = {1},
  journal = {npj Computational Materials},
  publisher = {Springer Science and Business Media LLC},
  author = {Ma,  Huan and Shang,  Honghui and Yang,  Jinlong},
  year = {2024},
  month = sep 
}

@article{DeepSolid,
  title = {Ab initio calculation of real solids via neural network ansatz},
  volume = {13},
  ISSN = {2041-1723},
  url = {http://dx.doi.org/10.1038/s41467-022-35627-1},
  DOI = {10.1038/s41467-022-35627-1},
  number = {1},
  journal = {Nature Communications},
  publisher = {Springer Science and Business Media LLC},
  author = {Li,  Xiang and Li,  Zhe and Chen,  Ji},
  year = {2022},
  month = dec 
}

@misc{Lv2025,
  doi = {10.48550/ARXIV.2507.02644},
  url = {https://arxiv.org/abs/2507.02644},
  author = {Gu,  Yuntian and Li,  Wenrui and Lin,  Heng and Zhan,  Bo and Li,  Ruichen and Huang,  Yifei and He,  Di and Wu,  Yantao and Xiang,  Tao and Qin,  Mingpu and Wang,  Liwei and Lv,  Dingshun},
  title = {Solving the Hubbard model with Neural Quantum States},
  publisher = {arXiv},
  year = {2025},
  copyright = {arXiv.org perpetual,  non-exclusive license}
}

@misc{Shang2025,
  doi = {10.48550/ARXIV.2509.25720},
  url = {https://arxiv.org/abs/2509.25720},
  author = {Ma,  Huan and Kan,  Bowen and Shang,  Honghui and Yang,  Jinlong},
  title = {Transformer-Based Neural Networks Backflow for Strongly Correlated Electronic Structure},
  publisher = {arXiv},
  year = {2025},
  copyright = {Creative Commons Attribution 4.0 International}
}

@misc{Di2025,
  doi = {10.48550/ARXIV.2509.09275},
  url = {https://arxiv.org/abs/2509.09275},
  author = {Zhang,  Lixing and Luo,  Di},
  title = {Neural Transformer Backflow for Solving Momentum-Resolved Ground States of Strongly Correlated Materials},
  publisher = {arXiv},
  year = {2025},
  copyright = {Creative Commons Attribution 4.0 International}
}

@misc{Javier2023,
  doi = {10.48550/ARXIV.2302.11588},
  url = {https://arxiv.org/abs/2302.11588},
  author = {Moreno,  Javier Robledo and Cohn,  Jeffrey and Sels,  Dries and Motta,  Mario},
  title = {Enhancing the Expressivity of Variational Neural,  and Hardware-Efficient Quantum States Through Orbital Rotations},
  publisher = {arXiv},
  year = {2023},
  copyright = {Creative Commons Attribution 4.0 International}
}

@inproceedings{AdamW,
title={Decoupled Weight Decay Regularization},
author={Ilya Loshchilov and Frank Hutter},
booktitle={International Conference on Learning Representations},
year={2019},
url={https://openreview.net/forum?id=Bkg6RiCqY7},
}

@article{Motta2017,
  title = {Towards the Solution of the Many-Electron Problem in Real Materials: Equation of State of the Hydrogen Chain with State-of-the-Art Many-Body Methods},
  volume = {7},
  ISSN = {2160-3308},
  url = {http://dx.doi.org/10.1103/PhysRevX.7.031059},
  DOI = {10.1103/physrevx.7.031059},
  number = {3},
  journal = {Physical Review X},
  publisher = {American Physical Society (APS)},
  author = {Motta,  Mario and Ceperley,  David M. and Chan,  Garnet Kin-Lic and Gomez,  John A. and Gull,  Emanuel and Guo,  Sheng and Jiménez-Hoyos,  Carlos A. and Lan,  Tran Nguyen and Li,  Jia and Ma,  Fengjie and Millis,  Andrew J. and Prokof’ev,  Nikolay V. and Ray,  Ushnish and Scuseria,  Gustavo E. and Sorella,  Sandro and Stoudenmire,  Edwin M. and Sun,  Qiming and Tupitsyn,  Igor S. and White,  Steven R. and Zgid,  Dominika and Zhang,  Shiwei},
  year = {2017},
  month = sep 
}

@misc{Loehr20025,
  doi = {10.48550/ARXIV.2510.26906},
  url = {https://arxiv.org/abs/2510.26906},
  author = {Loehr,  Kieran and Clark,  Bryan K.},
  title = {Enhancing Neural Network Backflow},
  publisher = {arXiv},
  year = {2025},
  copyright = {Creative Commons Attribution 4.0 International}
}

@article{Liu2025,
  title = {Efficient optimization of neural network backflow for ab initio quantum chemistry},
  volume = {112},
  ISSN = {2469-9969},
  url = {http://dx.doi.org/10.1103/thz7-lmdn},
  DOI = {10.1103/thz7-lmdn},
  number = {15},
  journal = {Physical Review B},
  publisher = {American Physical Society (APS)},
  author = {Liu,  An-Jun and Clark,  Bryan K.},
  year = {2025},
  month = oct 
}

@article{Chan2004,
  title = {State-of-the-art density matrix renormalization group and coupled cluster theory studies of the nitrogen binding curve},
  volume = {121},
  ISSN = {1089-7690},
  url = {http://dx.doi.org/10.1063/1.1783212},
  DOI = {10.1063/1.1783212},
  number = {13},
  journal = {The Journal of Chemical Physics},
  publisher = {AIP Publishing},
  author = {Chan,  Garnet Kin-Lic and Kállay,  Mihály and Gauss,  J\"{u}rgen},
  year = {2004},
  month = oct,
  pages = {6110–6116}
}

@article{Tubman2020,
  title = {Modern Approaches to Exact Diagonalization and Selected Configuration Interaction with the Adaptive Sampling CI Method},
  volume = {16},
  ISSN = {1549-9626},
  url = {http://dx.doi.org/10.1021/acs.jctc.8b00536},
  DOI = {10.1021/acs.jctc.8b00536},
  number = {4},
  journal = {Journal of Chemical Theory and Computation},
  publisher = {American Chemical Society (ACS)},
  author = {Tubman,  Norm M. and Freeman,  C. Daniel and Levine,  Daniel S. and Hait,  Diptarka and Head-Gordon,  Martin and Whaley,  K. Birgitta},
  year = {2020},
  month = mar,
  pages = {2139–2159}
}

@misc{Knitter2024,
  doi = {10.48550/ARXIV.2411.03900},
  url = {https://arxiv.org/abs/2411.03900},
  author = {Knitter,  Oliver and Zhao,  Dan and Stokes,  James and Ganahl,  Martin and Leichenauer,  Stefan and Veerapaneni,  Shravan},
  title = {Retentive Neural Quantum States: Efficient Ans\"{a}tze for Ab Initio Quantum Chemistry},
  publisher = {arXiv},
  year = {2024},
  copyright = {arXiv.org perpetual,  non-exclusive license}
}

@article{Tubman2016,
  title = {A deterministic alternative to the full configuration interaction quantum Monte Carlo method},
  volume = {145},
  ISSN = {1089-7690},
  url = {http://dx.doi.org/10.1063/1.4955109},
  DOI = {10.1063/1.4955109},
  number = {4},
  journal = {The Journal of Chemical Physics},
  publisher = {AIP Publishing},
  author = {Tubman,  Norm M. and Lee,  Joonho and Takeshita,  Tyler Y. and Head-Gordon,  Martin and Whaley,  K. Birgitta},
  year = {2016},
  month = jul 
}

@article{Holmes2016,
  title = {Heat-Bath Configuration Interaction: An Efficient Selected Configuration Interaction Algorithm Inspired by Heat-Bath Sampling},
  volume = {12},
  ISSN = {1549-9626},
  url = {http://dx.doi.org/10.1021/acs.jctc.6b00407},
  DOI = {10.1021/acs.jctc.6b00407},
  number = {8},
  journal = {Journal of Chemical Theory and Computation},
  publisher = {American Chemical Society (ACS)},
  author = {Holmes,  Adam A. and Tubman,  Norm M. and Umrigar,  C. J.},
  year = {2016},
  month = aug,
  pages = {3674–3680}
}

@article{Liu2024,
  title = {Neural network backflow for 
ab initio
 quantum chemistry},
  volume = {110},
  ISSN = {2469-9969},
  url = {http://dx.doi.org/10.1103/PhysRevB.110.115137},
  DOI = {10.1103/physrevb.110.115137},
  number = {11},
  journal = {Physical Review B},
  publisher = {American Physical Society (APS)},
  author = {Liu,  An-Jun and Clark,  Bryan K.},
  year = {2024},
  month = sep 
}

@article{Li2024,
  title = {Improved optimization for the neural-network quantum states and tests on the chromium dimer},
  volume = {160},
  ISSN = {1089-7690},
  url = {http://dx.doi.org/10.1063/5.0214150},
  DOI = {10.1063/5.0214150},
  number = {23},
  journal = {The Journal of Chemical Physics},
  publisher = {AIP Publishing},
  author = {Li,  Xiang and Huang,  Jia-Cheng and Zhang,  Guang-Ze and Li,  Hao-En and Shen,  Zhu-Ping and Zhao,  Chen and Li,  Jun and Hu,  Han-Shi},
  year = {2024},
  month = jun 
}

@misc{Malyshev2024,
  doi = {10.48550/ARXIV.2408.07625},
  url = {https://arxiv.org/abs/2408.07625},
  author = {Malyshev,  Aleksei and Schmitt,  Markus and Lvovsky,  A. I.},
  title = {Neural Quantum States and Peaked Molecular Wave Functions: Curse or Blessing?},
  publisher = {arXiv},
  year = {2024},
  copyright = {Creative Commons Attribution Non Commercial Share Alike 4.0 International}
}

@article{ForemanMackey2013,
  title = {emcee: The MCMC Hammer},
  volume = {125},
  ISSN = {1538-3873},
  url = {http://dx.doi.org/10.1086/670067},
  DOI = {10.1086/670067},
  number = {925},
  journal = {Publications of the Astronomical Society of the Pacific},
  publisher = {IOP Publishing},
  author = {Foreman-Mackey,  Daniel and Hogg,  David W. and Lang,  Dustin and Goodman,  Jonathan},
  year = {2013},
  month = mar,
  pages = {306–312}
}

@misc{Zhuo2022,
  doi = {10.48550/ARXIV.2212.06835},
  url = {https://arxiv.org/abs/2212.06835},
  author = {Chen,  Zhuo and Luo,  Di and Hu,  Kaiwen and Clark,  Bryan K.},
  title = {Simulating 2+1D Lattice Quantum Electrodynamics at Finite Density with Neural Flow Wavefunctions},
  publisher = {arXiv},
  year = {2022},
  copyright = {arXiv.org perpetual,  non-exclusive license}
}

@article{White1999,
  title = {Ab initio quantum chemistry using the density matrix renormalization group},
  volume = {110},
  ISSN = {1089-7690},
  url = {http://dx.doi.org/10.1063/1.478295},
  DOI = {10.1063/1.478295},
  number = {9},
  journal = {The Journal of Chemical Physics},
  publisher = {AIP Publishing},
  author = {White,  Steven R. and Martin,  Richard L.},
  year = {1999},
  month = mar,
  pages = {4127–4130}
}

@article{White1992,
  title = {Density matrix formulation for quantum renormalization groups},
  author = {White, Steven R.},
  journal = {Phys. Rev. Lett.},
  volume = {69},
  issue = {19},
  pages = {2863--2866},
  numpages = {0},
  year = {1992},
  month = {Nov},
  publisher = {American Physical Society},
  doi = {10.1103/PhysRevLett.69.2863},
  url = {https://link.aps.org/doi/10.1103/PhysRevLett.69.2863}
}

@article{Coester1960,
  title = {Short-range correlations in nuclear wave functions},
  volume = {17},
  ISSN = {0029-5582},
  url = {http://dx.doi.org/10.1016/0029-5582(60)90140-1},
  DOI = {10.1016/0029-5582(60)90140-1},
  journal = {Nuclear Physics},
  publisher = {Elsevier BV},
  author = {Coester,  F. and K\"{u}mmel,  H.},
  year = {1960},
  month = jun,
  pages = {477–485}
}

@article{Bytautas2009,
	author = {Laimutis Bytautas and Klaus Ruedenberg},
	doi = {https://doi.org/10.1016/j.chemphys.2008.11.021},
	issn = {0301-0104},
	journal = {Chemical Physics},
	note = {Moving Frontiers in Quantum Chemistry:},
	number = {1},
	pages = {64-75},
	title = {A priori identification of configurational deadwood},
	url = {https://www.sciencedirect.com/science/article/pii/S0301010408005314},
	volume = {356},
	year = {2009}}

@article{Anderson2018,
	author = {James S.M. Anderson and Farnaz Heidar-Zadeh and Paul W. Ayers},
	doi = {https://doi.org/10.1016/j.comptc.2018.08.017},
	issn = {2210-271X},
	journal = {Computational and Theoretical Chemistry},
	pages = {66-77},
	title = {Breaking the curse of dimension for the electronic Schr{\"o}dinger equation with functional analysis},
	url = {https://www.sciencedirect.com/science/article/pii/S2210271X18304250},
	volume = {1142},
	year = {2018}}

@article{Hermann2020,
	author = {Hermann, Jan and Sch{\"a}tzle, Zeno and No{\'e}, Frank},
	da = {2020/10/01},
	doi = {10.1038/s41557-020-0544-y},
	id = {Hermann2020},
	isbn = {1755-4349},
	journal = {Nature Chemistry},
	number = {10},
	pages = {891--897},
	title = {Deep-neural-network solution of the electronic Schr{\"o}dinger equation},
	ty = {JOUR},
	url = {https://doi.org/10.1038/s41557-020-0544-y},
	volume = {12},
	year = {2020}}

@inproceedings{Wu2023,
  series = {SC ’23},
  title = {NNQS-Transformer: an Efficient and Scalable Neural Network Quantum States Approach for Ab initio Quantum Chemistry},
  url = {http://dx.doi.org/10.1145/3581784.3607061},
  DOI = {10.1145/3581784.3607061},
  booktitle = {Proc. Int. Conf. High Perform. Comput. Netw. Storage Anal.},
  publisher = {ACM},
  author = {Wu,  Yangjun and Guo,  Chu and Fan,  Yi and Zhou,  Pengyu and Shang,  Honghui},
  year = {2023},
  month = nov,
  pages = {1–13},
  collection = {SC ’23}
}

@article{zejun2023,
  title = {Unifying view of fermionic neural network quantum states: From neural network backflow to hidden fermion determinant states},
  author = {Liu, Zejun and Clark, Bryan K.},
  journal = {Phys. Rev. B},
  volume = {110},
  issue = {11},
  pages = {115124},
  numpages = {17},
  year = {2024},
  month = {Sep},
  publisher = {American Physical Society},
  doi = {10.1103/PhysRevB.110.115124},
  url = {https://link.aps.org/doi/10.1103/PhysRevB.110.115124}
}

@article{pyscf,
  title = {PySCF: the Python‐based simulations of chemistry framework},
  volume = {8},
  ISSN = {1759-0884},
  url = {http://dx.doi.org/10.1002/wcms.1340},
  DOI = {10.1002/wcms.1340},
  number = {1},
  journal = {WIREs Computational Molecular Science},
  publisher = {Wiley},
  author = {Sun,  Qiming and Berkelbach,  Timothy C. and Blunt,  Nick S. and Booth,  George H. and Guo,  Sheng and Li,  Zhendong and Liu,  Junzi and McClain,  James D. and Sayfutyarova,  Elvira R. and Sharma,  Sandeep and Wouters,  Sebastian and Chan,  Garnet Kin‐Lic},
  year = {2017},
  month = sep 
}

@misc{Malyshev2023,
  doi = {10.48550/ARXIV.2310.04166},
  url = {https://arxiv.org/abs/2310.04166},
  author = {Malyshev,  Aleksei and Arrazola,  Juan Miguel and Lvovsky,  A. I.},
  title = {Autoregressive Neural Quantum States with Quantum Number Symmetries},
  publisher = {arXiv},
  year = {2023},
  copyright = {Creative Commons Attribution Non Commercial Share Alike 4.0 International}
}

@article{Sharma2017,
  title = {Semistochastic Heat-Bath Configuration Interaction Method: Selected Configuration Interaction with Semistochastic Perturbation Theory},
  volume = {13},
  ISSN = {1549-9626},
  url = {http://dx.doi.org/10.1021/acs.jctc.6b01028},
  DOI = {10.1021/acs.jctc.6b01028},
  number = {4},
  journal = {Journal of Chemical Theory and Computation},
  publisher = {American Chemical Society (ACS)},
  author = {Sharma,  Sandeep and Holmes,  Adam A. and Jeanmairet,  Guillaume and Alavi,  Ali and Umrigar,  C. J.},
  year = {2017},
  month = mar,
  pages = {1595–1604}
}

@misc{Tubman2018,
  doi = {10.48550/ARXIV.1808.02049},
  url = {https://arxiv.org/abs/1808.02049},
  author = {Tubman,  Norm M. and Levine,  Daniel S. and Hait,  Diptarka and Head-Gordon,  Martin and Whaley,  K. Birgitta},
  title = {An efficient deterministic perturbation theory for selected configuration interaction methods},
  publisher = {arXiv},
  year = {2018},
  copyright = {arXiv.org perpetual,  non-exclusive license}
}

@article{Cleland2012,
  title = {Taming the First-Row Diatomics: A Full Configuration Interaction Quantum Monte Carlo Study},
  volume = {8},
  ISSN = {1549-9626},
  url = {http://dx.doi.org/10.1021/ct300504f},
  DOI = {10.1021/ct300504f},
  number = {11},
  journal = {Journal of Chemical Theory and Computation},
  publisher = {American Chemical Society (ACS)},
  author = {Cleland,  Deidre and Booth,  George H. and Overy,  Catherine and Alavi,  Ali},
  year = {2012},
  month = oct,
  pages = {4138–4152}
}

@article{Pfau2020,
  title = {Ab initio
 solution of the many-electron Schr\"{o}dinger equation with deep neural networks},
  volume = {2},
  ISSN = {2643-1564},
  url = {http://dx.doi.org/10.1103/PhysRevResearch.2.033429},
  DOI = {10.1103/physrevresearch.2.033429},
  number = {3},
  journal = {Physical Review Research},
  publisher = {American Physical Society (APS)},
  author = {Pfau,  David and Spencer,  James S. and Matthews,  Alexander G. D. G. and Foulkes,  W. M. C.},
  year = {2020},
  month = sep 
}

@misc{Shang2023,
  doi = {10.48550/ARXIV.2307.09343},
  url = {https://arxiv.org/abs/2307.09343},
  author = {Shang,  Honghui and Guo,  Chu and Wu,  Yangjun and Li,  Zhenyu and Yang,  Jinlong},
  title = {Solving Schr\"{o}dinger Equation with a Language Model},
  publisher = {arXiv},
  year = {2023},
  copyright = {Creative Commons Attribution 4.0 International}
}

@article{Zhao2023,
	author = {Tianchen Zhao and James Stokes and Shravan Veerapaneni},
	doi = {10.1088/2632-2153/acdb2f},
	journal = {Machine Learning: Science and Technology},
	month = {jun},
	number = {2},
	pages = {025034},
	publisher = {IOP Publishing},
	title = {Scalable neural quantum states architecture for quantum chemistry},
	url = {https://dx.doi.org/10.1088/2632-2153/acdb2f},
	volume = {4},
	year = {2023}}

@article{Li2023,
	author = {Li, Xiang and Huang, Jia-Cheng and Zhang, Guang-Ze and Li, Hao-En and Cao, Chang-Su and Lv, Dingshun and Hu, Han-Shi},
	da = {2023/11/28},
	date = {2023/11/28},
	doi = {10.1021/acs.jctc.3c00831},
	isbn = {1549-9618},
	journal = {Journal of Chemical Theory and Computation},
	journal1 = {J. Chem. Theory Comput.},
	m3 = {doi: 10.1021/acs.jctc.3c00831},
	month = {11},
	number = {22},
	pages = {8156--8165},
	publisher = {American Chemical Society},
	title = {A Nonstochastic Optimization Algorithm for Neural-Network Quantum States},
	ty = {JOUR},
	url = {https://doi.org/10.1021/acs.jctc.3c00831},
	volume = {19},
	year = {2023},
	year1 = {2023}}

@article{Di2019,
	author = {Luo, Di and Clark, Bryan K.},
	doi = {10.1103/PhysRevLett.122.226401},
	issue = {22},
	journal = {Phys. Rev. Lett.},
	month = {Jun},
	numpages = {6},
	pages = {226401},
	publisher = {American Physical Society},
	title = {Backflow Transformations via Neural Networks for Quantum Many-Body Wave Functions},
	url = {https://link.aps.org/doi/10.1103/PhysRevLett.122.226401},
	volume = {122},
	year = {2019}}

\end{document}